\documentclass{jfm}
\usepackage{graphicx}
\usepackage{epstopdf, epsfig}
\usepackage{caption}
\usepackage{amsmath, amssymb}
\usepackage{mathtools}
\usepackage{subcaption}
\usepackage{soul}
\usepackage{tikz,siunitx}

\usepackage{cleveref}
\usepackage{siunitx}
\usepackage{lineno}
\usepackage{placeins}
\usepackage{marginnote}
\newcommand{\Rei}{\text{Re}_i}

\newcommand{\pof}{Physics of Fluids Group, Max Planck UT Center for Complex Fluid Dynamics, MESA+ Institute and J.M. Burgers Centre for Fluid Dynamics, University of Twente, P.O. Box 217, 7500 AE Enschede, The Netherlands}
\newcommand{\bundes}{Institut f\"ur Str\"omungsmechanik und Aerodynamik, Universit\"at der Bundeswehr M\"unchen, Werner-Heisenberg-Weg 39, 85577 Neubiberg, Germany}

\shorttitle{TC flow with rough surfaces}
\shortauthor{P. Berghout et al.}

\title{Characterizing the turbulent drag properties of rough surfaces with a Taylor--Couette setup }

\author{Pieter Berghout\aff{1}\corresp{\email{p.berghout@utwente.nl}},
Pim A. Bullee\aff{1,2}, Thomas Fuchs\aff{3}, \\ Sven Scharnowski\aff{3}, Christian J. K\"ahler\aff{3}, Daniel Chung\aff{4}, \\ Detlef Lohse\aff{1,5} and Sander G. Huisman\aff{1}\corresp{\email{s.g.huisman@utwente.nl}}}

\affiliation{\aff{1}\pof \aff{2}Soft Matter, Fluidics and Interfaces, MESA+ Research Institute, University of Twente, P.O. Box 217, 7500AE Enschede, The Netherlands, \aff{3}\bundes, \aff{4} Department of Mechanical Engineering, University of Melbourne, Victoria 3010, Australia, \aff{5} Max Planck Institute for Dynamics and Self-Organization, Am Fa\ss berg 17, 37077 G\"ottingen, Germany}
\begin{document}

\maketitle

\begin{abstract}
Wall-roughness induces extra drag in wall-bounded turbulent flows. Mapping any given roughness geometry to its fluid dynamic behaviour has been hampered by the lack of accurate and direct measurements of skin-friction drag. Here the Taylor--Couette (TC) system provides an opportunity as it is a closed system and allows to directly and reliably measure the skin-friction. However, the wall-curvature potentially complicates the connection between the wall friction and the wall roughness characteristics. Here we investigate the effects of a hydrodynamically fully rough surface on highly turbulent, inner cylinder rotating, TC flow. We carry out particle image velocimetry (PIV) measurements in the Twente Turbulent Taylor--Couette (T$^3$C) facility with radius ratio $\eta = 0.720$ at Reynolds numbers in the range of $4.6\times 10^5 < \Rei < 1.77\times10^6$, with water as working fluid. The inner cylinder is covered with P36 grit sandpaper, and the outer cylinder remains smooth and stationary.
We find that the effects of a hydrodynamically fully rough surface on TC turbulence, where the roughness height $k$ is three orders of magnitude smaller than the Obukhov curvature length $L_c$ (which characterizes the effects of curvature on the turbulent flow, see Berghout \textit{et al.} arXiv: 2003.03294, 2020), are similar to those effects of a fully rough surface on a flat plate turbulent boundary layer (BL). Hence, the value of the equivalent sand grain height $k_s$, that characterizes the drag properties of a rough surface, is similar to those found for comparable sandpaper surfaces in a flat plate BL.
Next, we obtain the dependence of the torque (skin-friction drag) on the Reynolds number for given wall roughness, characterized by $k_s$, and find agreement with the experimental results within $5\%$. Our findings demonstrate that global torque measurements in the TC facility are well suited to reliably deduce wall drag properties for any rough surface.
\end{abstract}

\section{Introduction}\label{sec:introduction}
\subsection{Turbulent boundary layers over fully rough walls}
The transport of a fluid over a solid body or the transport of a solid body through a fluid is always hindered by friction forces acting on the interface between the solid and the fluid. Ideally, the solid surface is smooth, and the drag force is a purely viscous force. In nature and engineering applications, however, solid surfaces are nearly always rough. This means that in addition to a modified viscous force, the roughness also results in a pressure contribution to the drag force (`pressure drag'), and consequently, an increase in the total drag force (the so-called `drag penalty'). The contribution of the pressure drag to the total friction drag at the surface grows with increasing roughness height. Ultimately, when the pressure drag  dominates, the surface is called \textit{hydrodynamically fully rough}.

Due to the obvious interest in reducing the drag penalty, substantial research has been carried out to investigate the effects of rough surfaces on wall-bounded turbulent flows \citep{jim04, fla10, chu21}. The key effect thereof is a downward shift (by $\Delta u^+$) of the mean streamwise velocity ($u^+$) in the overlap (or logarithmic) region of the turbulent BL \citep{cla54,ham54}. This shift can be considered as a direct measure of the drag penalty. The mean velocity profile for a rough wall in the overlap region is given by the Prandtl--von K\'arm\'an profile for smooth walls, minus this shift~\citep{pop00}
    \begin{equation}\label{eq:log_smooth}
    u^+ = \frac{1}{\kappa}\log y^+ + A - \Delta u^+,
    \end{equation}
where $y^+$ is the wall-normal distance and the von K\'{a}rm\'{a}n constant $\kappa \approx 0.40$ and $A\approx5.0$ are extracted from experimental or numerical data. The superscript `$+$' as usual indicates a normalization with the viscous velocity scale $u_\tau = \sqrt{\tau_w/\rho}$ and the viscous length scale $\delta_\nu = \nu/u_\tau$, where $\tau_w$ is the wall shear stress, $\rho$ the fluid density and $\nu$ is the kinematic viscosity. For a fully rough surface, it can be derived from dimensional arguments that the velocity shift $\Delta u^+$ depends logarithmically on the roughness height $k^+$, see e.g. \citep{rau91,pop00}. The so-called fully rough asymptote of the roughness function is given by
    \begin{equation}\label{eq:fra}
    \Delta u^+ = \frac{1}{\kappa} \log k_s^+ +A -B,
    \end{equation}
where $B\approx 8.5$ is the Nikuradse constant. The equivalent sand grain roughness height $k_s^+$ is obtained by fitting, such that the velocity shift of any fully rough surface collapses with the velocity shift of sand grains in turbulent pipe flow, that historically grew to be the reference case \citep{nik33}. Hence, the key objective in research of wall bounded turbulent flows over rough surfaces is to relate the statistics of a rough surface to the value of $k_s$, which characterizes the roughness \citep{for17}.

\subsection{Taylor--Couette flow}
TC flow --- the flow between two coaxial, independently, rotating cylinders --- is a canonical system in turbulence \citep{tay23, gro16}. Since the domain is closed in all directions, global balances can be derived and monitored, giving room for extensive comparison between theory, experiments and simulations. Moreover, the torque (corresponding to the skin-friction) can be measured accurately and directly~\citep{gil12,hui14}, in contrast to measurements of skin-friction in open systems.

The forcing strength of the system is quantified by the ratio of the centrifugal force and the viscous force, i.e. the Taylor number
    \begin{equation}\label{eq:Ta}
    Ta = \frac{1}{4}\left(\frac{1+\eta}{2\sqrt{\eta}}\right)^4 \frac{(r_o - r_i)^2(r_i+r_o)^2(\omega_i-\omega_o)^2}{\nu^2}.
    \end{equation}
Here $\eta$ is the geometric measure of curvature, namely the ratio $r_i/r_o$ of the radii of the cylinders. The subscripts $i$ and $o$ indicate inner cylinder and outer cylinder, respectively. The angular velocity is denoted by $\omega$, and $\nu$ is the kinematic viscosity.

The global response of the system is expressed as the Nusselt number $Nu_\omega$, which is the ratio between the angular velocity flux $J^\omega$ in radial direction and its laminar counterpart $J^\omega_\text{lam}$~\citep{eck07b}, as
    \begin{equation}\label{eq:Nuw}
    Nu_\omega = \frac{J^\omega}{J^\omega_\text{lam}} = \frac{ r^3(\left<u_r\omega\right>_{A(r),t} - \nu \partial_r \left<\omega\right>_{A(r),t}) }  {2\nu r_i^2 r_o^2(\omega_i - \omega_o)/(r_o^2 - r_i^2)}.
    \end{equation}
Here, $\langle \cdot \rangle_{A(r),t}$ denotes averaging over the cylinder surface $A(r)$ and over time $t$.
The Nusselt number $Nu_\omega$ is related to the torque $\mathcal{T}$ required to drive the inner cylinder. In non-dimensional form the torque can be expressed as
    \begin{equation}\label{eq:G}
    G = \frac{\mathcal{T}}{2\pi L \rho \nu^2} = Nu_\omega \frac{J_\text{lam}^\omega}{\nu^2}.
    \end{equation}
From here onwards, we assume inner cylinder rotation only, hence $\omega_o = 0$, as this corresponds to our experiments where we kept the (smooth) outer cylinder stationary at all times.

The torque is directly related to the wall shear stress $\tau_w = \mathcal{T}/(2 \pi r_i^2 L)$.
As commonly used in other canonical systems (e.g. the flat plate BL), we define the friction factor $C_f$ as \citep{lat92a}:
    \begin{equation}
    \label{eq:cf}
    C_f = \frac{2\pi \tau_{w,i}}{\rho d^2  \omega_i^2}= 2\pi Nu_\omega J_{lam}^\omega (\nu Re_i)^{-2},
\end{equation}
where $Re_i=r_i\omega_i d/\nu$ and $d=r_o-r_i$. This relation allows for straightforward comparison with other canonical wall-bounded flows like pipe flow, channel flow, flow over a flat plate, etc.

The turbulent flow in the TC setup is strongly influenced by the curvature of its bounding walls, i.e. the cylinders that drive the flow. This distinguishes turbulent TC flow from turbulent flows in other canonical systems. \cite{bra69} realized that the effects of curvature on a turbulent BL are very similar to the effects of buoyancy stratification on a turbulent BL \citep{obu71}. In analogy to the Obukhov length \citep{obu71,mon75}, he derived a length scale that separates the curved BL in a region where the effects of shear dominate (i.e. production of turbulence is dominated by shear production), and a region further away from the wall where curvature effects dominate (i.e. the production of turbulence is dominated by curvature). For smooth wall TC turbulence, this `curvature Obukhov length' is well approximated by~\citep{ber20b}
    \begin{equation}\label{eq:lc}
    L_{c,s} = \frac{u_\tau}{\kappa \omega_i},
    \end{equation}
where shear dominates for $0.20L_{c,s} \lesssim y$, shear and curvature effects are both significant for $0.20L_{c,s} \lesssim y \lesssim 0.65$, and curvature effects dominate at $0.65L_{c,s} \lesssim y$. Using data of the mean velocity profiles from PIV in turbulent TC flow~\citep{hui13, vee16} and direct numerical simulations (DNS) \citep{ost15b}, the mean angular logarithmic velocity profile in the region of the turbulent BL where curvature effects are important was recently obtained for smooth wall TC flow~\citep{ber20b}. By employing a matching argument between the velocity profiles of the turbulent BL and the bulk region, following the work of \cite{che19}, an analytical expression for $Nu(Ta)$ was derived~\citep{ber20b}.

The effects of irregular boundaries (extended transverse bars in the `obstacle regime', as referred to by \cite{jim04}) on turbulent TC flow was previously investigated by means of experiments \citep{cad97, ber03, zhu18, ver18} and DNS \citep{zhu17, zhu18}. Here the ratio $k/d$ between the height of the bars and the gap width $d=r_o-r_i$ was as large as $k/d = 0.05$ or even $0.1$. Later \cite{ber19} numerically studied the effects of sand grain roughness ($k/d = \numrange{0.019}{0.087}$) on the turbulent TC velocity profiles, and found similar transitionally rough behaviour as the sand grain roughness of \cite{nik33} in turbulent pipe flow. However, we note that both the experimental and computational studies in TC flow suffered from limited scale separation between the roughness scale $k$ and the gap width $d$.

In this paper, we study the effects of a hydrodynamically fully rough inner cylinder on the turbulent wall-bounded flow, with small roughness $k / d = 0.014$, where $k \equiv 6k_\sigma$, and $k_\sigma$ is the standard deviation of the roughness elevation. In particular, we keep $k$ smaller than the curvature Obukhov length $L_c$ (see (\ref{eq:blr})), namely $k/L_c = \numrange{0.078}{0.090}$. We will demonstrate that in order to study the effects of roughness on a turbulent flow in TC, $k\ll d$ is not enough. Rather, $k\ll L_c$ must also hold, to ensure that effects related to the streamwise curved geometry are not influencing the effects of the roughness.

Hence, we hypothesize that effects of roughness in TC turbulence (where $k\ll L_c$) are similar to the effects of roughness in other canonical systems without streamwise curvature. Thus global measurements in the (closed) TC facility can be employed to characterize drag properties of the rough surface. The outer cylinder remains smooth, to allow for optical access of the velocity profiles.

The paper is organised as follows: In \S~\ref{sec:exp_methods}, we describe the experimental methods. We then (\S~\ref{sec:scale_separation}) discuss the relevant dynamical length scales in the experiment, and elaborate on the different regions in the BL where turbulent production is dominated by shear effects, and where effects related to the streamwise curvature of the setup play a role. We also comment on the scale separation and show that the roughness mainly affects the inertial shear dominated regime, and hence, effects from the streamwise curved geometry of the TC flow do not modify the velocity shift. In \S~\ref{sec:mean} we use the mean velocity profiles of the inner cylinder boundary layer, to show that apart from the shift, the velocity profiles for  rough and smooth inner cylinders are the same. We use this in \S~\ref{sec:fra} to calculate the angular velocity shift $\Delta \omega^+$, from which the equivalent sand grain roughness height is determined in \S~\ref{sec:equisan}. In \S~\ref{sec:angmom} we demonstrate that the bulk region of the flow is of constant angular momentum, which is used in \S~\ref{sec:nu_ta} to obtain a relation between the torque (skin-friction drag) and the Reynolds number for given surface roughness $k_s$, in agreement with our experimental results. The paper ends with a summary, conclusions, and an outlook (\S~\ref{sec:sum_con}).

\section{Experimental setup and methods}\label{sec:exp_methods}
    \begin{figure}
    \centering
    \includegraphics[width=0.7\linewidth]{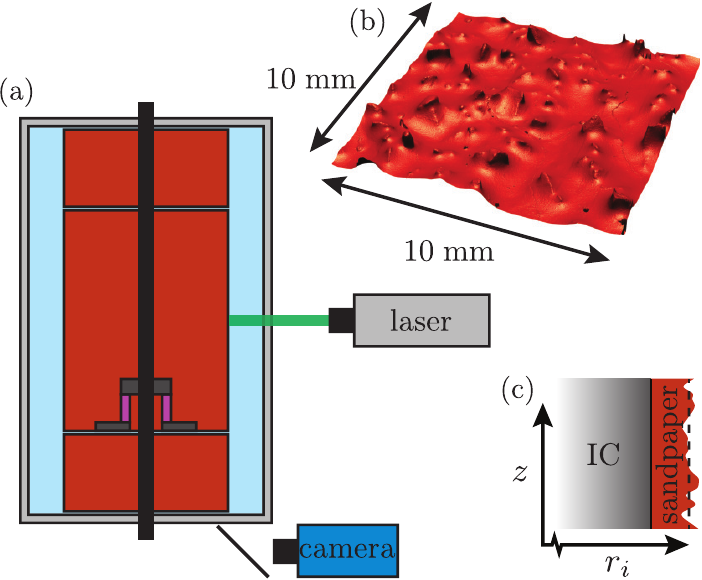}
        \caption{(a) Cross section of the TC geometry. The tracer particles are illuminated by a $\SI{1}{\mm}$ thick horizontal laser light-sheet from the right. The light scattered by the tracers is imaged from the bottom through a mirror. The torque sensor only measured the torque of the middle part of the inner cylinder. (b) 3D visualization of the confocal scan of the used sandpaper. (c) Cross-section of the inner cylinder with the sandpaper attached to the surface.}
    \label{fig:setup}
    \end{figure}
\subsection{Experimental setup}
The experiments were performed in the Twente Turbulent Taylor--Couette (T$^3$C) facility \citep{gil11a}, with water as working liquid. We used a fully rough inner cylinder with an outer radius of $r_i = \SI{201.2}{\mm}$, and a transparent outer cylinder with an inner radius of $r_o = \SI{279.4}{\mm}$. This gives a radius ratio of $\eta=0.720$ and a gap width $d= \SI{78.2}{\mm}$. The cylinders have a height of $L=\SI{927}{\mm}$ and an aspect ratio of $\Gamma = L/d = 11.9$. For inner cylinder rotation only (the outer cylinder is stationary), the Reynolds number is defined with the velocity of the inner cylinder and the gap width $d$ as
\begin{equation}\label{eq:Rei}
     Re_i = \frac{\omega_i r_i d}{\nu}.
\end{equation}
Using the viscous velocity $u_\tau$ obtained from measurements of the torque, the friction Reynolds number is defined as
\begin{equation}\label{eq:Retau}
     Re_\tau = \frac{u_\tau (d/2)}{\nu}.
\end{equation}
The roughness used was P36 grit sandpaper (\texttt{VSM}, ceramic industrial-grade), that was fixed to the entire surface of the inner cylinder using double-sided adhesive tape (\texttt{tesa 51970}). We define the characteristic length scale of the roughness as $k \equiv 6k_\sigma \approx \SI{1.07}{\mm}$ (corresponding to the $99.8\%$ interval of the height), where $k_\sigma$ is the standard deviation of the local roughness height $h(x,y)$ (quantified using confocal microscopy~\citep{bak20} over a square part of the roughness sandpaper with width \SI{25}{\mm}), and $k/d = 0.014$.

\subsection{Experimental procedure}
We performed seven experiments with different rotation rates of the inner cylinder, see table \ref{table_param}. During all these experiments, the torque $\mathcal{T}$ that is required to drive the inner cylinder at fixed rotational velocity was measured constantly. The hollow reaction torque sensor that connects the drive shaft to the middle section of the inner cylinder is indicated in figure~\ref{fig:setup}(a). By only measuring the torque on the middle section, possible end-plate effects are eliminated~\citep{gil12}. During the torque measurements, PIV was used to obtain the velocity field in the gap. To quantify the reproducibility of our torque measurements, we compared the torque data that were captured during the PIV experiments with three separate torque measurements thereafter. We find a spread in $\mathcal{T}$ smaller than $4\%$ for all cases. These direct and reproducible measurements of the torque (friction) have an accuracy that is comparable to the measurement accuracy of wall shear stress in flat plate BLs, by means of a drag balance \citep{baa16}.

\begin{table}
\centering
\begin{tabular}{c c c c c c c c c c c}
$Ta [\times 10^{12}]$ & $Re_i [\times 10^{6}]$  & $Nu_\omega$  & $C_f [\times 10^{-3}]$ & $Re_\tau[\times 10^3]$ & $L_c^+[\times 10^3]$ &  $k_{\sigma}^+$  \\
\hline
$0.31$ & $0.46$  & $312$   & $2.21$ & $7.6$ & $2.6$ &  $34$  \\
$0.57$ & $0.62$  & $403$   & $2.13$ & $10.0$ & $3.3$ &  $44$  \\
$0.92$ & $0.78$  & $513$   & $2.14$ & $12.7$ & $4.1$ &  $57$  \\
$1.47$ & $0.99$  & $643$   & $2.12$ & $16.0$ & $5.0$ &  $72$  \\
$2.18$ & $1.20$  & $784$   & $2.12$ & $19.5$ & $6.0$ &  $87$  \\
$3.55$ & $1.54$  & $998$   & $2.11$ & $24.9$ & $7.6$ &  $111$ \\
$4.71$ & $1.77$  & $1137$  & $2.09$ & $28.5$ & $8.5$ &  $127$ \\
$6.15$ & $2.00$  & $653$   & $1.07$ & $23.1$ & $6.9$ &  $0$
\end{tabular}
\caption{Control parameters, global response and relevant length scales, measured during the PIV measurements. $Ta$ or $Re_i$ characterize the driving of the system. $Nu_\omega$ is the dimensionless angular velocity flux, $C_f$ the friction factor, $L_c^+$ the curvature Obukhov length as defined in equation (\ref{eq:blr}) and $k_\sigma^+$ is the standard deviation of the sandpaper roughness, see \cite{bak20}. The final row presents the values correponding to the smooth wall measurement of \cite{hui13}}\label{table_param}
\end{table}
For the PIV measurements, fluorescent polymer tracer particles (\texttt{Dantec FPP-RhB-10} with diameters from $\SI{1}{\micro\metre}$ to $\SI{20}{\micro\metre}$) were added to the working fluid. A horizontal laser sheet of about $\SI{1}{\mm}$ in thickness illuminated tracer particles in the working liquid at mid-height, through the transparent outer cylinder. The laser sheet was created using a frequency doubled \texttt{Quantel EverGreen \SI{200}{\milli\joule}} laser. The fluorescent light emerging from the tracer particles was imaged from below, through a window placed in the bottom plate of the apparatus. For this, a $45^\circ$ mirror was positioned under the bottom plate as drawn schematically in figure~\ref{fig:setup}. The camera was a high-resolution sCMOS camera (\texttt{LaVision PCO.edge}), with a resolution of $2560\,\text{px} \times 2160\,\text{px}$ and a pixel size of \SI{6.5}{\micro\metre}. A \SI{100}{\mm} focal length objective (\texttt{Zeiss Makro Planar}, \SI{100}{mm}) was used, giving an optical magnification of $0.17$.

For each rotational velocity of the inner cylinder, $10^4$ image pairs were acquired at a recording frequency equal to the rotation rate. The mean velocity distribution in the horizontal plane was computed using single-pixel ensemble correlation \citep{kahler2006,kah12a}. The spatial resolution is \SI{50}{\micro\metre}, leading to about 1600 independent measurement points in the radial direction, evenly spread over the entire gap. From the correlation function (obtained for every pixel) one can directly extract the standard deviation of the velocity, by integrating the probability density function $\sigma(u) = \int \nolimits_{-\infty}^\infty (u-\langle u \rangle)^2 \,\,\text{PDF}(u)\,du$, see \cite{sch12b}. This ensures that all turbulent scales are included in the standard deviation, as opposed to regular PIV analysis. The velocity profiles were smoothed using a Gaussian filter with a standard deviation of $\sigma \approx \SI{0.5}{\mm}$.

\section{Curvature effects, the mean velocity profile and scale separation}\label{sec:scale_separation}
\subsection{The relative effects of curvature and shear}\label{sec:curv_strat}
To characterize and quantify the relative effects of shear and curvature in TC turbulence, we study the ratio $S$ of turbulence production by shear and curvature related effects \citep{bra69, tow76,ber20b}
     \begin{equation}\label{eq:S}
        S^{-1} = \frac{\overline{u'_\theta u'_r}\frac{d}{d r} U}
             {\frac{1}{r}\overline{u'_\theta u'_r} U} = \frac{1}{\omega}\frac{dU}{dr} ,
    \end{equation}
where $u'_\theta$ and $u'_r$ are the azimuthal and radial velocity fluctuations, respectively, and $\overline{u'_\theta u'_r}$ is the Reynolds stress. The mean azimuthal velocity is denoted by $U$, and $\omega = U/r$ is the mean angular velocity. The curvature Obukhov length $L_c$ defined in equation~(\ref{eq:lc}) for a smooth wall marks the transition from a region where the production of turbulence is dominated by shear ($y<0.20L_c$), to a region where it is affected by curvature ($y>0.20L_c$). Hence, by this definition, for $S=1$ we have $y^+=L_c^+$. The definition from equation~(\ref{eq:lc}) builds on the existence of a shear logarithmic region, where the gradient of the mean angular velocity is $\frac{d}{d r} U = u_\tau/(\kappa y)$. The angular velocity scale for rough walls is approximated as $\omega=\omega_i + \Delta \omega$. Thus the generic curvature Obukhov length $L_c$ for smooth and rough walls can be defined with the inner cylinder rotation rate $\omega_i$, and the wall-shear stress $\tau_w$ only, similar to equation~\ref{eq:lc}, but now for a rough wall,
\begin{equation}
\label{eq:blr}
L_c = \frac{u_{\tau}}{\kappa (\omega_i + \Delta \omega)},
\end{equation}
so that $L_c^+(\Delta \omega^+ = 0)=L_{c,s}^+$.

\begin{figure}
\centering
\begin{subfigure}{.50\textwidth}
\centering
\includegraphics[width=0.97\linewidth]{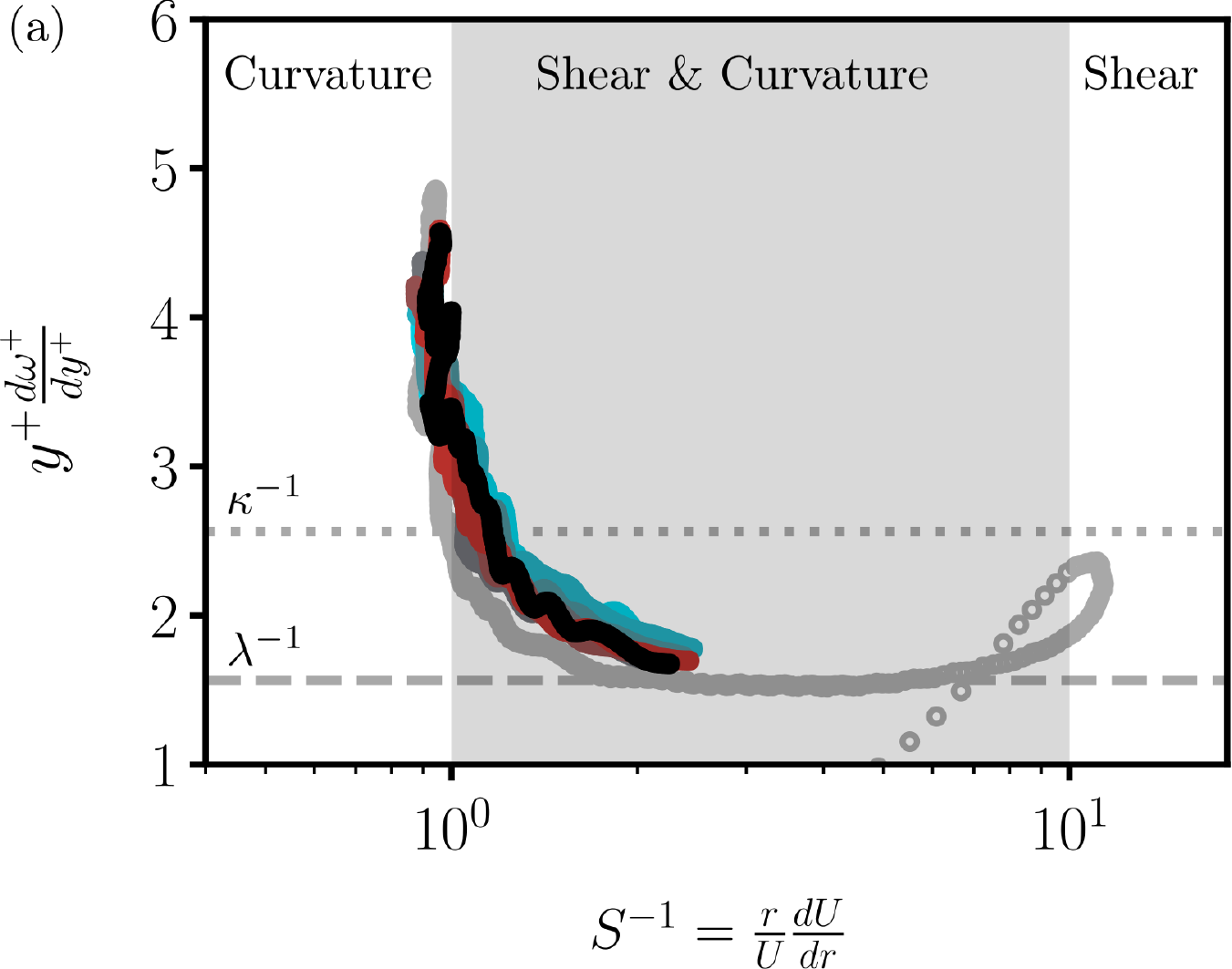}
\end{subfigure}%
\begin{subfigure}{.50\textwidth}
\centering
\includegraphics[width=0.97\linewidth]{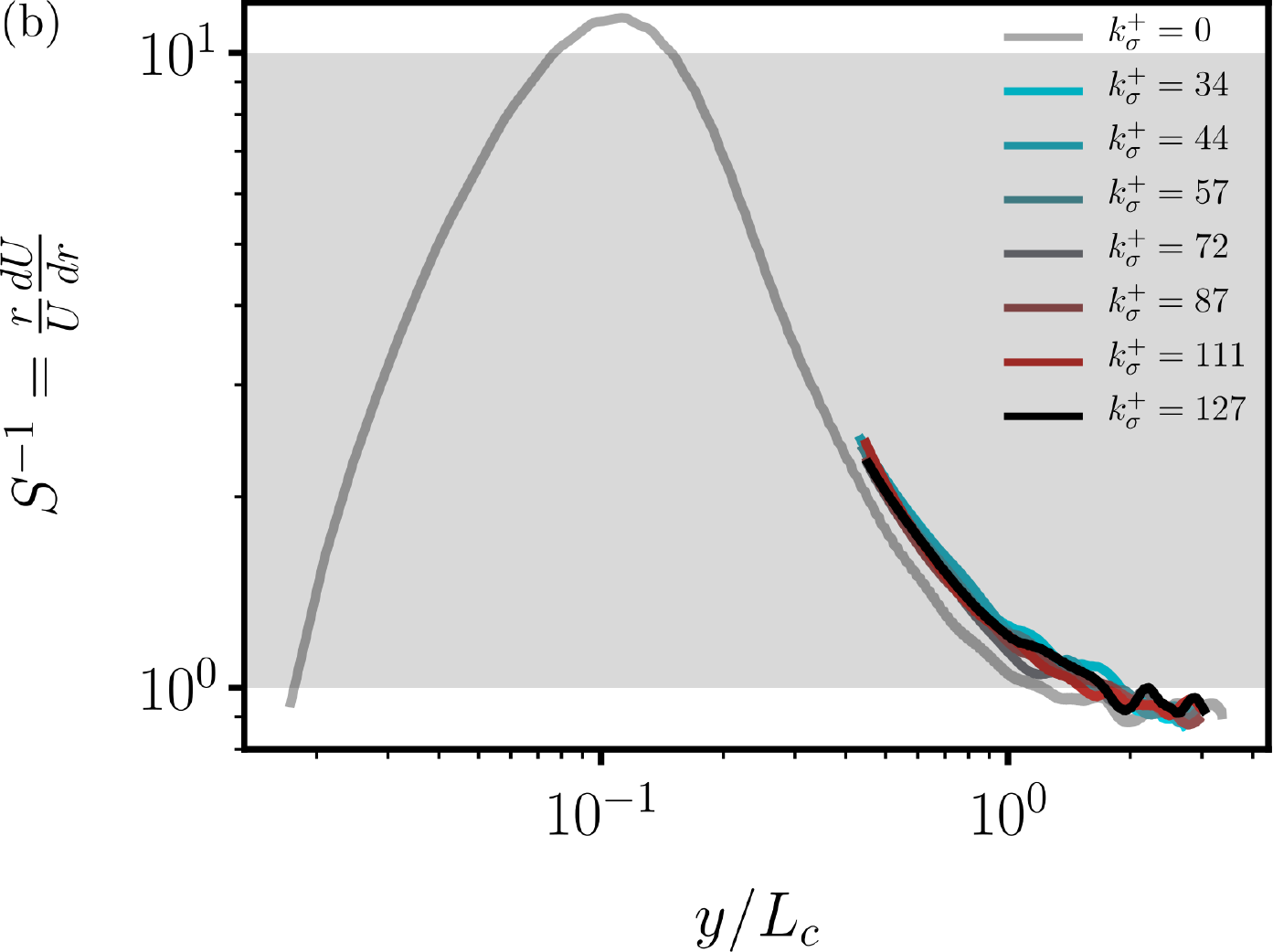}
\end{subfigure}%
\caption{(a) Compensated gradient of the mean angular velocity profile versus the ratio $S$ between the turbulence production by shear and that by curvature, see equation (\ref{eq:S}). Dotted and dashed lines represent the slope of the logarithmic velocity profile of the shear and the curvature dominated regimes, $\kappa^{-1}$ and $\lambda^{-1}$, respectively. (b) Ratio $S$ versus the wall-normal distance shifted with the wall offset, $y/L_c = (r-r_i-2k_\sigma)/L_c$, where $2k_\sigma$ is the approximated wall offset of the sandpaper. Coloured lines are calculated from the PIV data of the rough wall cases. The grey line ($k_\sigma^+=0$) is the smooth wall profile at $Ta=6.2\times 10^{12}$ ($Re_\tau=23093$), obtained from \cite{hui13}.}
\label{fig:S_full}
\end{figure}
Figure \ref{fig:S_full}(a) presents the gradient of the mean angular velocity profile versus $S$, calculated from the PIV results. We find fair collapse of the velocity gradients of smooth (grey) and rough (coloured) wall profiles. When the effects of curvature are negligible $S \ge\mathcal{O}(10)$, the gradient of the velocity profile approaches $\kappa^{-1} \approx 2.5$. This occurs in a very small region close to the wall, where we cannot measure due to the presence of the sandpaper roughness. For the rough and smooth wall velocity profiles, we find that the gradient approaches $\lambda^{-1}$ in the region where curvature and shear affect the flow. For $S\le 1$, curvature effects dominate the flow, and a constant angular momentum region (i.e. the bulk flow) sets in \citep{ber20b}.

\subsection{The mean angular velocity profile}\label{sec:meanangvel}
\begin{figure}
\centering
\includegraphics[width=\textwidth]{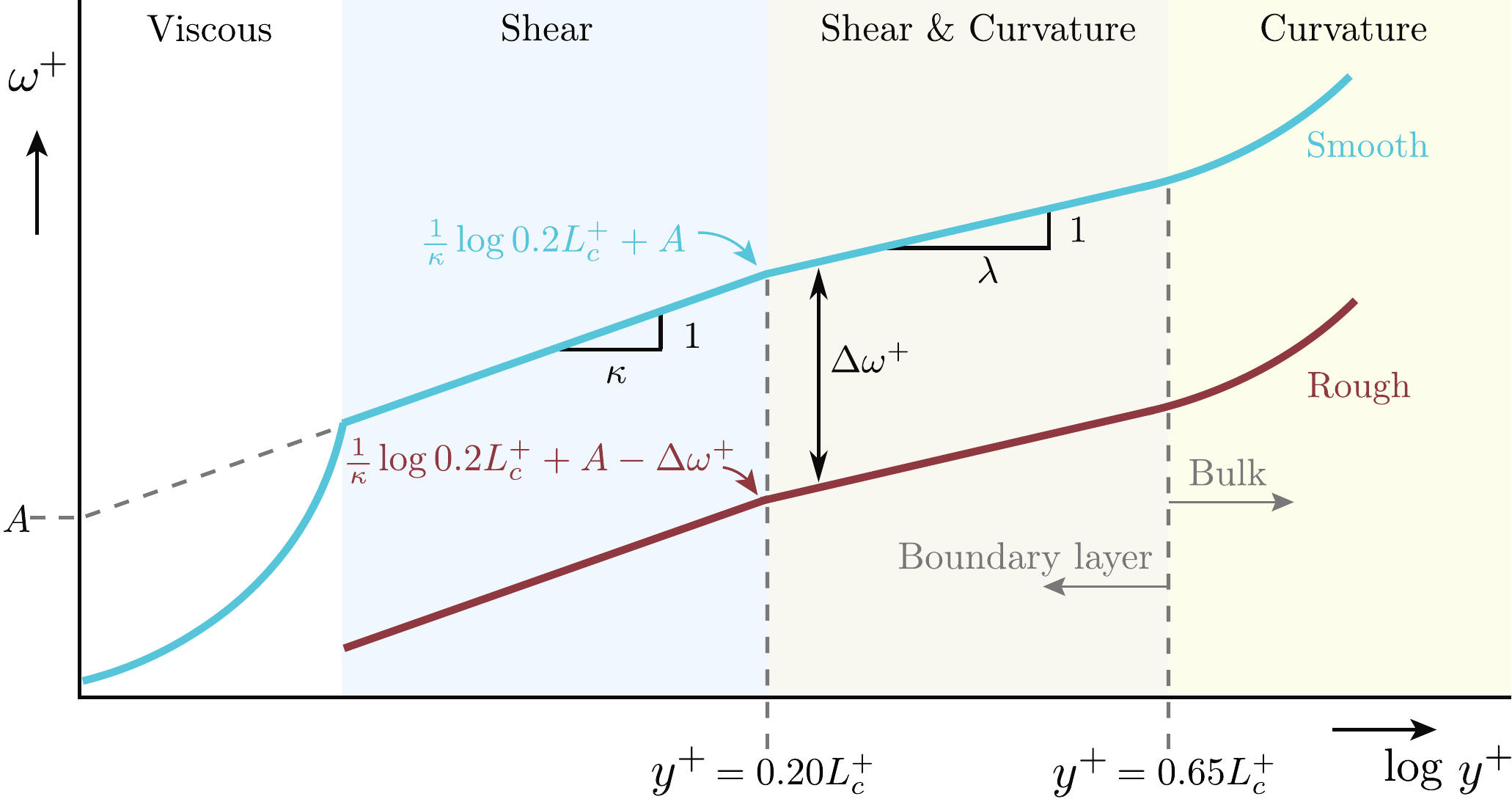}
\caption{Schematic of the various regions in smooth and IC-rough turbulent TC flow at matched $L_c^+$. The height $y^+ = 0.65L_c^+$ is defined as the location where the logarithmic profile with slope $\lambda^{-1}$ ends and the constant angular momentum region of the bulk velocity starts.}
\label{fig:ill_velprof}
\end{figure}
Figure \ref{fig:S_full}(a) shows that the curvature and shear affected region of the BL contains a constant gradient ($=\lambda^{-1}$) of the mean angular velocity. From this observation, \cite{ber20b} obtained an equation of the mean angular velocity in the shear and curvature affected region in the BL (in short `curvature log').


The offset of the logarithmic velocity profile (with slope $\lambda^{-1}$) in the curvature and shear affected region, as indicated in figure~\ref{fig:ill_velprof}, is a function of the wall normal location where curvature related effects impact the flow. From PIV results, the exact location was found to be $y^+=0.20L_c^+$, with $L_c^+$ defined in (\ref{eq:blr}). Therefore, the offset is $\kappa^{-1}\log 0.2L_c^+ + A$, where $A=5.0$ is the offset of the logarithmic velocity profile in the shear affected region \citep{pop00}. The transition in the logarithmic velocity profile from the shear affected region to the curvature and shear affected region at $y^+=0.20L_c^+$ is not sharp but gradual. To account for this, we introduce a constant $C_{bl}$ that connects the logarithmic velocity profiles of both regions. \cite{ber20b} found that $A+C_{bl}+\left( \frac{1}{\kappa} - \frac{1}{\lambda} \right)\log (0.2) =1.0$, for the inner cylinder, and the mean angular velocity equation, above $y^+=0.20L_c^+$ as
\begin{equation}
    \label{eq:smooth_rough}
    \omega^+ = \frac{1}{\kappa} \log 0.2L_c^+ + A + C_{bl} + \frac{1}{\lambda} \log \frac{y^+}{0.2L_c^+} = \frac{1}{\lambda}\log y^+ + \left( \frac{1}{\kappa}-\frac{1}{\lambda}\right)\log L_{c}^+ + 1.0, 
\end{equation}
with $C_{bl}=-3.30$.
The transition from the curvature and shear affected region to the constant angular momentum region occurs at $y^+=0.65L_c^+$. This height we take as our definition of the boundary layer height, above which is the bulk region of constant angular momentum.

\subsection{Scale separation}\label{sec:scale_sep}
Key to the understanding of the effects of roughness in TC turbulence is the concept of scale separation. To illustrate this, in figure \ref{fig:S_full}(b) we plot $S$ versus the wall-normal distance $y/L_c = (r-r_i-2k_\sigma)/L_c$. We note that the wall offset $2k_\sigma^+$ of the rough wall is an approximation. As a reference, we also plot the smooth wall profile (grey) at $Ta=6.2\times 10^{12}$ \citep{hui13}, together with the rough wall profiles (colors).

Table \ref{table_param} presents the relevant dynamical length scales in the experiments: namely, $Re_\tau$, $L_c^+$ and $k_\sigma^+$. The friction Reynolds number $Re_\tau$ from equation (\ref{eq:Retau}) gives the ratio of the largest dynamical length scale in the TC setup to the viscous length scale $\delta_\nu$. $Re_\tau$ is of the same order as in the smooth wall TC experiments by \cite{hui13}, where it was $Re_\tau = \numrange{488}{23093}$, comparable to the rough BL experiments by \cite{squ16}, where $Re_\tau = \numrange{2890}{29900}$.

The roughness scale in our experiments is much larger than the viscous length scale $\delta_\nu$, i.e. $k_\sigma^+ = \numrange{34}{127} \gg 1$, and thus pressure drag dominates over viscous drag. For the flat plate BL experiments of \cite{squ16} in the fully rough regime, we estimate that $k_\sigma^+ = \numrange{9}{12}$ is required, based on the data for which $\Delta U^+ > 8.0$ in \cite{squ16}. Hence, we are confident that we are indeed far in the fully rough regime. We also find that the roughness sublayer height $\approx 3k_\sigma^+$ is smaller than the outer bound of the shear dominated logarithmic region, $\approx 0.2L_c^+$. For the lowest roughness we have $3k_\sigma^+/0.2L_c^+=0.40$, and for the highest roughness it is $3k_\sigma^+/0.2L_c^+=0.20$. This separation of length scales allows for a region where the logarithmic velocity profile can form. For example in the smooth wall experiments of~\cite{hui13} such a profile was found between $50 \le y^+ \le 600$ for comparable $Ta$. We finally find that the outer bound of the curvature dominated logarithmic region $L_c^+$ is smaller than the outer length scale $Re_\tau$, so that $0.65L_c^+/Re_\tau \approx 0.33$. For $y^+>0.65L_c^+$ the curvature dominated bulk, constant angular momentum region, forms. The occurrence and extent of this constant angular momentum region depends on the radius ratio $\eta$, i.e. $L_c^+/Re_\tau$ depends on $\eta$.

Table 1 suggests that the roughness only affects the inertial region where curvature effects are negligible. Hence, we expect that the velocity shift of that region is similar to that of identical sandpaper in a flat plate turbulent BL. In other words, we would expect a fully rough asymptote with slope $\kappa^{-1}$ and a similar value of $k_s$ as we would measure for identical sandpaper in a flat plate turbulent BL.

\section{Mean velocity profiles of the inner cylinder boundary layer}\label{sec:mean}
\begin{figure}
\centering
\begin{subfigure}{.50\textwidth}
\centering
\includegraphics[width=0.97\linewidth]{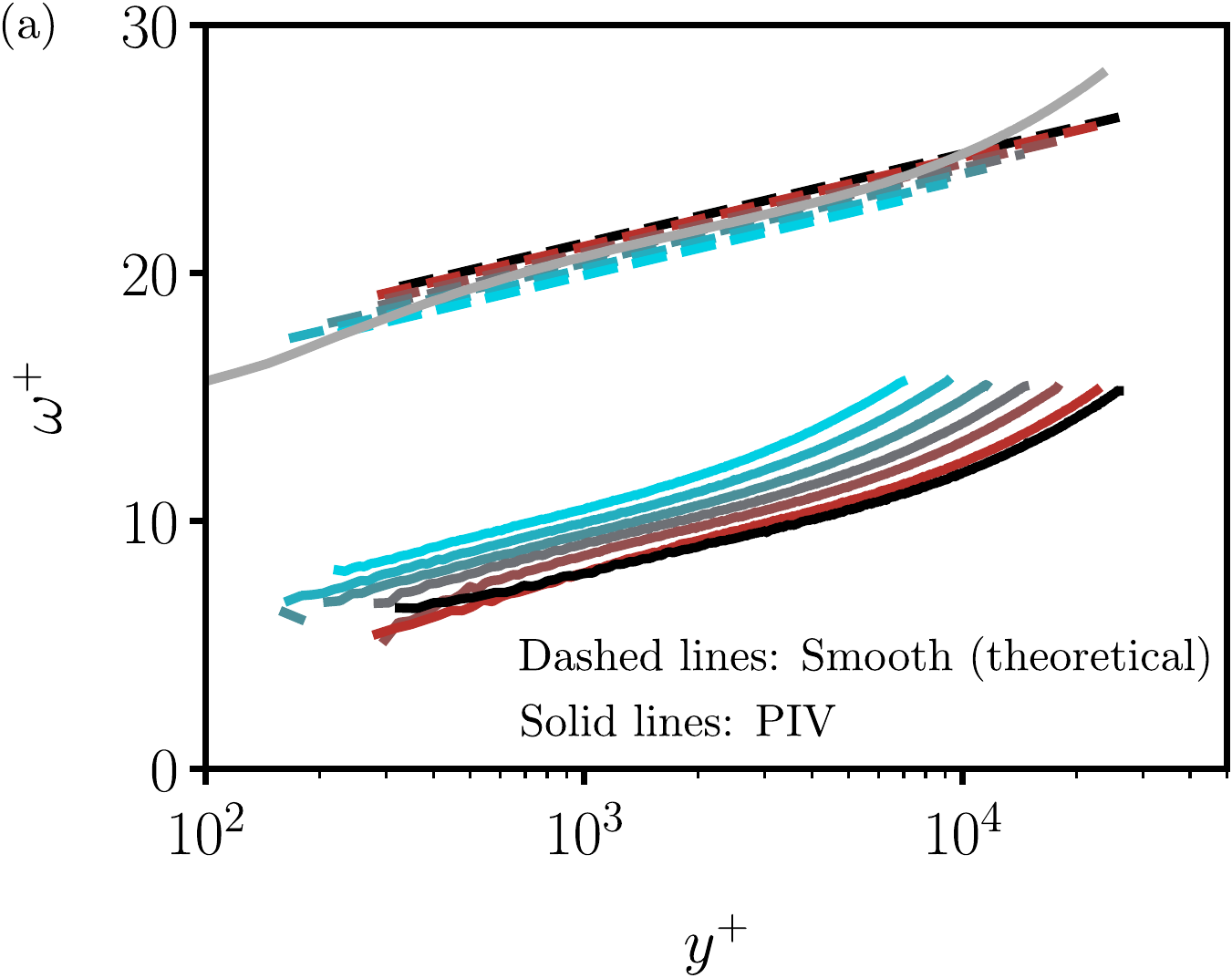}
\end{subfigure}%
\begin{subfigure}{.50\textwidth}
\centering
\includegraphics[width=0.97\linewidth]{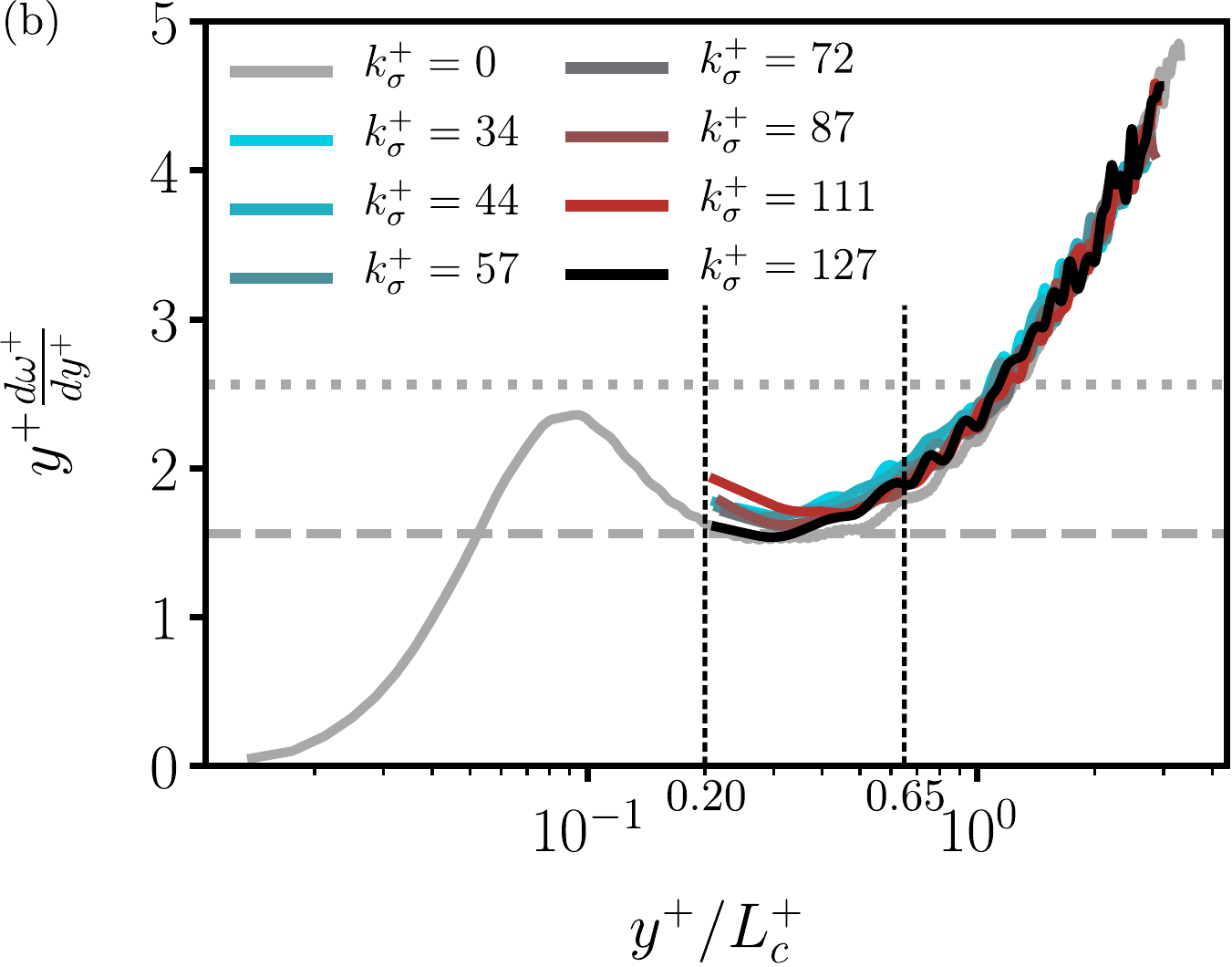}
\end{subfigure}%
\caption{(a) Mean angular velocity $\omega^+$ versus the wall-normal distance $y^+$. The solid lines are the measured rough wall profiles. The dashed lines represent the theoretical smooth wall reference profiles (colors are the same), calculated from equation (\ref{eq:smooth_rough}) and at matching $L_c^+$ and $Re_\tau$ with the rough wall profiles. (b) The compensated gradient of the rough wall profiles in (a), where the wall-normal distance is normalized with the curvature length $L_c^+$. The colors are the same in both figures. The grey line is the smooth wall profile at $Ta=6.2\times10^{12}$, obtained from \cite{hui13}. The dashed horizontal line represents the slope $\lambda^{-1}$ of the logarithmic velocity profile in the region where turbulence production is dominated by curvature effects.
The dotted horizontal line represents the slope $\kappa^{-1}$ of the logarithmic velocity profile in the region where turbulence production is dominated by shear.}
\label{fig:omega}
\end{figure}
In this paper we will show the angular velocity profile $\omega^+(y^+)$ rather than the azimuthal velocity profile $u^+(y^+)$, as it is $\omega^+(y^+)$ which is expected, given the arguments based on the Navier--Stokes equations, to have a logarithmic profile \citep{gro14}. 

Figure \ref{fig:omega}(a) shows the angular velocity profiles over the rough wall $\omega^+ = \langle \omega_i - \omega(r)\rangle_t/\omega_{\tau}$, with $\omega_{\tau}=u_{\tau}/r_i$, versus the wall-normal coordinate $y^+$.
In this and the next section, we focus our analysis on the mean velocity profiles of the inner cylinder BL, hence $u_\tau = u_{\tau,i}$ throughout. In \S \ref{sec:angmom} we will report on the bulk profiles. We refer to \cite{ber20b} for an analysis of the smooth velocity profiles of the outer cylinder BL.

Figure \ref{fig:omega}(a) shows that with increasing roughness the rough wall profiles are increasingly shifted downwards, as expected. More importantly, we find from the diagnostic function $y^+\frac{d\omega^+}{dy^+}$ (a useful representation of the gradients \citep{pop00}) in figure \ref{fig:omega}(b), that the slope $\lambda^{-1}$ of the curvature dominated logarithmic region is the same for rough wall TC turbulence as for smooth wall TC turbulence (grey line). Unfortunately, we could not resolve the very thin spatial region where a shear dominated logarithmic was found by \cite{hui13}, as the roughness peaks obstruct the view for the PIV very close to the wall.

For a rough wall, $\Delta \omega^+$ is a function of the equivalent sand grain roughness height $k_s^+$ and the curvature length $L_c$, so that $\Delta \omega^+ (k_s^+, \frac{k_s}{L_c})$. When $k_s \ll L_c$, the angular velocity shift only depends on $k_s$, and the shift becomes $\Delta \omega^+ (k_s^+)$. Since the inner cylinder rotates, the plus sign in the denominator of equation~(\ref{eq:blr}) is connected with the increase of angular fluid velocity in the inner cylinder BL due to the roughness. When we normalize the wall-normal distance with $L_c^+$, we expect the transition from a curvature logarithmic velocity profile to the constant angular momentum bulk velocity profile to occur at $y^+/L_c^+ = 0.65$.
In figure \ref{fig:omega}(b) we find a fair collapse of both smooth and rough wall profiles in wall-normal direction, when normalized with curvature length $L_c^+$.

\section{The fully rough asymptote}\label{sec:fra}
From the observation that both smooth and rough wall velocity profiles possess the same slope $\lambda^{-1}$ of the curvature logarithmic region (figure \ref{fig:omega}b) we proceed to calculate the angular velocity shift $\Delta \omega^+$. Due to the roughness the angular velocity profiles in the shear logarithmic region are shifted, as discussed in \S \ref{sec:curv_strat} and illustrated in figure \ref{fig:ill_velprof}. This shift remains also in the curvature logarithmic region, where we will now quantify it. The offset of that region scales with $ \frac{1}{\kappa} \log(0.20L_{c}^+)+A$ (figure \ref{fig:ill_velprof}). Hence, it is imperative to calculate the angular velocity shift from the smooth wall velocity profile at matching $L_c^+$.

\begin{figure}
\centering
\begin{subfigure}{.50\textwidth}
\centering
\includegraphics[width=0.97\linewidth]{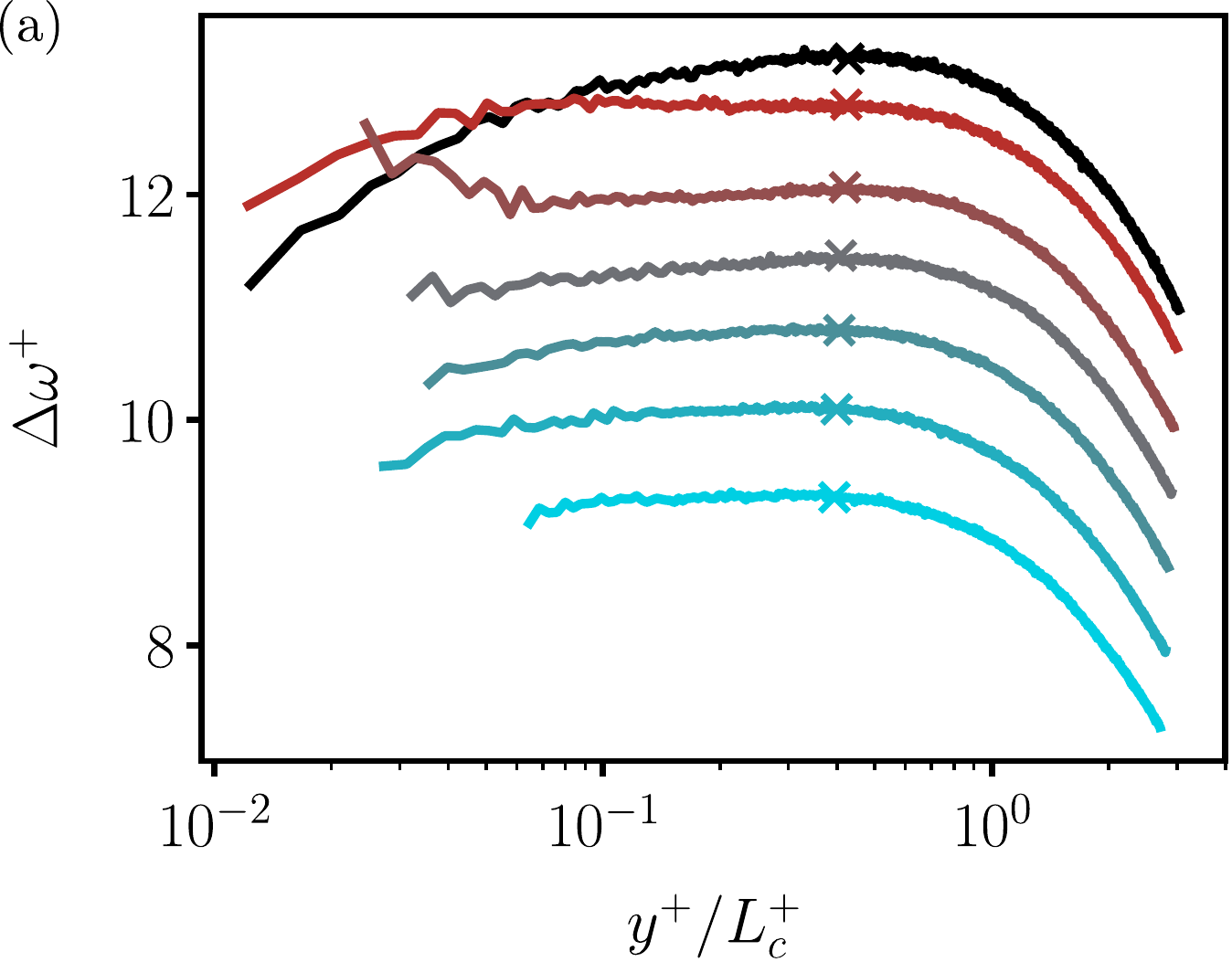}
\end{subfigure}%
\begin{subfigure}{.50\textwidth}
\centering
\includegraphics[width=0.97\linewidth]{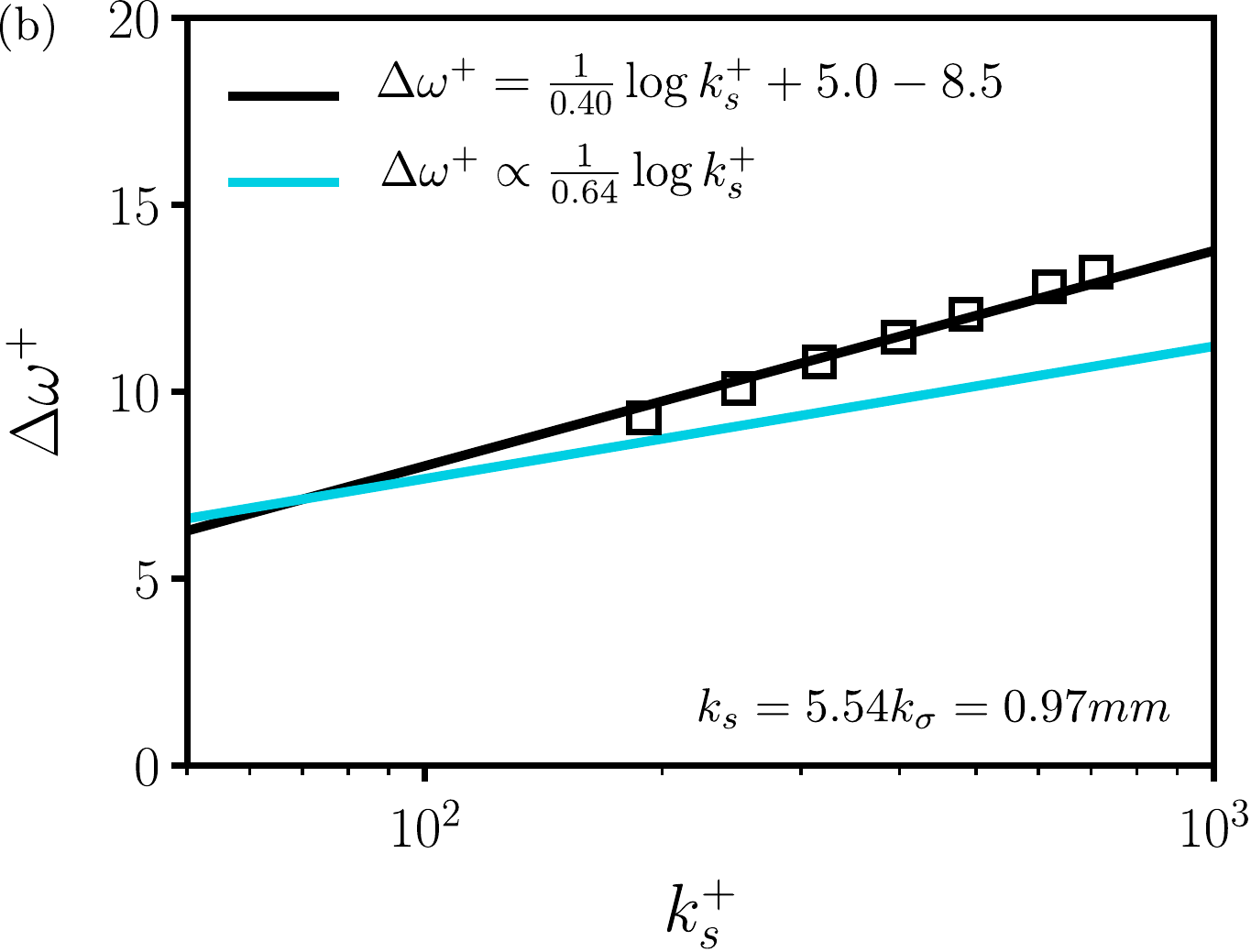}
\end{subfigure}%
\caption{$\Delta \omega^+$ of the rough wall profiles with respect to the reference smooth wall profiles. (a) Velocity shift versus the wall-normal distance $y^+/L_c^+$ (colors the same as figure \ref{fig:omega}). (b) The velocity shift $\Delta \omega^+$, crosses in (a), versus the equivalent sand grain height $k_s^+$. Black symbols are the experimental values. The solid black line is the fully rough asymptote of \cite{nik33}, equation~(\ref{eq:fr_asymptote}). The solid blue line is an illustration of the curvature fully rough asymptote, with slope $\lambda^{-1}$ and arbitrary vertical shift.}
\label{fig:fra}
\end{figure}

Figure \ref{fig:omega}(a) shows these smooth wall profiles (dashed), where the colors match the respective rough wall cases, and both $L_{c}^+$ and $Re_{\tau}$ are matched.
The velocity shift $\Delta \omega^+ (y^+)$ from the theoretical smooth wall profile, equation~(\ref{eq:smooth_rough}), is plotted in figure \ref{fig:fra}(a). The horizontal plateaus confirm the similarity of the slopes of the velocity profiles. We extract $\Delta \omega^+$ at $y^+ \approx 0.4 L_c^+$ and plot the shift versus the roughness height in figure \ref{fig:fra}(b).
When we fit a function of the form $\Delta \omega^+ = \frac{1}{a}\log k^+ + b$ through all seven data points, we obtain $a=0.34\pm 0.02$, to within $15\%$ of the von K\'arm\'an constant $\kappa \approx 0.38$, the slope of the shear dominated logarithmic profile. This confirms our hypothesis, as discussed in section \ref{sec:curv_strat}, that the fully rough asymptote for $\delta_\nu \ll k < L_c^+$ has slope $\kappa^{-1}$. For reference, this is much higher than $\lambda^{-1}\approx 0.64^{-1}$, blue line in figure \ref{fig:fra}(b).

To obtain a measure of the equivalent sandgrain roughness height $k_s$, we fit the data points to the fully rough asymptote of \cite{nik33}
\begin{equation}\label{eq:fr_asymptote}
\Delta \omega^+ (k_s^+) = \frac{1}{\kappa}\log k_s^+ + 5.0-8.5,
\end{equation}
and obtain $k_s=5.54k_\sigma=\SI{0.97}{\mm}$, for $\kappa\approx 0.40$. For reference, the typical grain size is estimated by $6k_\sigma=\SI{1.05}{\mm}$ \cite{bak20}.

\section{The equivalent sand grain roughness height}\label{sec:equisan}

\begin{figure}
\centering
\includegraphics[width=1.0\linewidth]{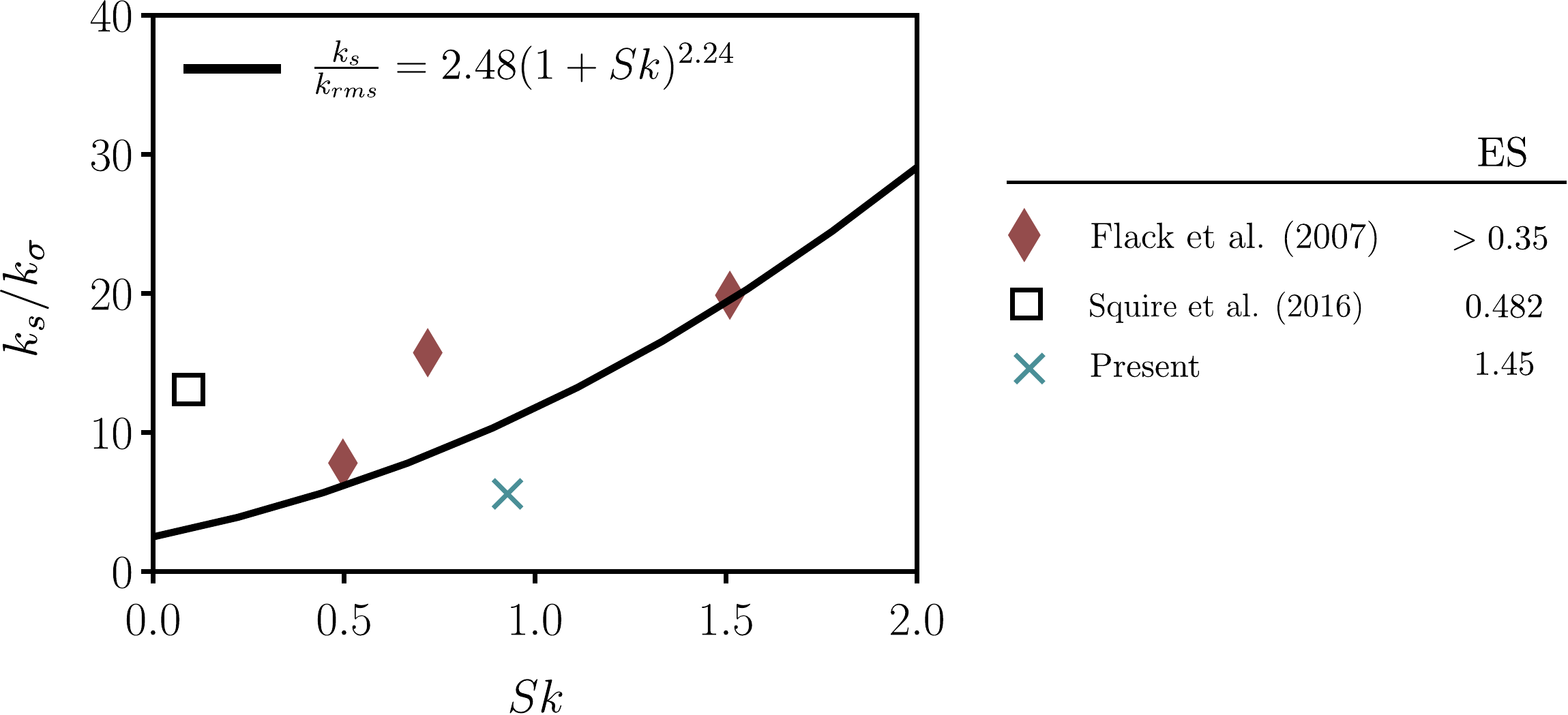}
\caption{Relationship between the equivalent sand grain roughness height divided by the root-mean-square height $k_s/k_{\sigma}$, and the skewness parameter $Sk$ of different sandpaper surfaces. The solid black line is the empirical correlation for $Sk>0$ from \cite{fla20}. Data from turbulent boundary layer flow using grit (12, 24, and 80) sandpaper \citep{fla07}, of which the surface statistics are listed in \cite{fla10}, turbulent boundary layer using grit 36 \citep{squ16} and turbulent TC flow using grit 36 (present). ES is the effective slope defined as $|\overline{\frac{dk}{d(r\theta)}}|$ \citep{nap08}.}
\label{fig:ks_final}
\end{figure}
The hypothesis in this research, postulated in \S \ref{sec:curv_strat}, is that the fully rough asymptote in TC turbulence with $\delta_\nu \ll k < L_c$ is the same (or very similar) to the fully rough asymptote in flows without streamwise curvature. We have already demonstrated in \S \ref{sec:fra} that the slope $\kappa^{-1}$ of the fully rough asymptote is indeed (almost) the same. This leaves us with a comparison of the value of $k_s$, between TC turbulence and canonical systems without streamwise curvature.

In literature, we have found two reports on turbulent flows over sandpaper roughness: the work of \cite{squ16}, employing 36 grit sandpaper in a turbulent BL, and \cite{fla07} who employed (12-, 24-, and 80-) grit sandpaper in turbulent BL flow. In the rough wall TC experiments reported here we used grit 36 sandpaper. However, it is essential to realize that sandpaper is not only defined by the grit size. Other statistics, like the skewness (an important parameter \citep{for17}, which is 0.93 here, and only 0.09 in \cite{squ16}), do vary with manufacturing methods. We have tried to use the very same sandpaper type (SP40F, \texttt{Awuko Abrasives}) as \cite{squ16}. Unfortunately, the sandpaper turned out to be not waterproof, and detached from the inner cylinder. We then applied new water resistant sandpaper (\texttt{VSM}, P36 grit ceramic industrial grade), with different surface roughness statistics.

To compare the drag property of the sandpaper surfaces in TC, to the respective sandpaper surfaces in literature, we plot the relationship between $k_s$ and the root-mean-square and skewness in figure \ref{fig:ks_final}. The surface properties of the sandpaper surface from \cite{fla07} are taken from \cite{fla10}. The solid black line is the empirical correlation from \cite{fla20}.
We find that the relation between $k_s$ and the Skewness $Sk$ and the root-mean-square height $k_{rms}$ of sandpaper used in our rough wall TC experiments is consistent with the empirical trend given for the sandpaper used in rough wall turbulent BL flow analysis. Whether, the deviation originates from the difference between TC and canonical systems without curvature, or originates from the different surface statistics (e.g. the ES for the present surface is higher, indicating a denser surface), remains to be resolved.

\section{The constant angular momentum region in the bulk}\label{sec:angmom}
\begin{figure}
\centering
\begin{subfigure}{.50\textwidth}
\centering
\includegraphics[width=0.97\linewidth]{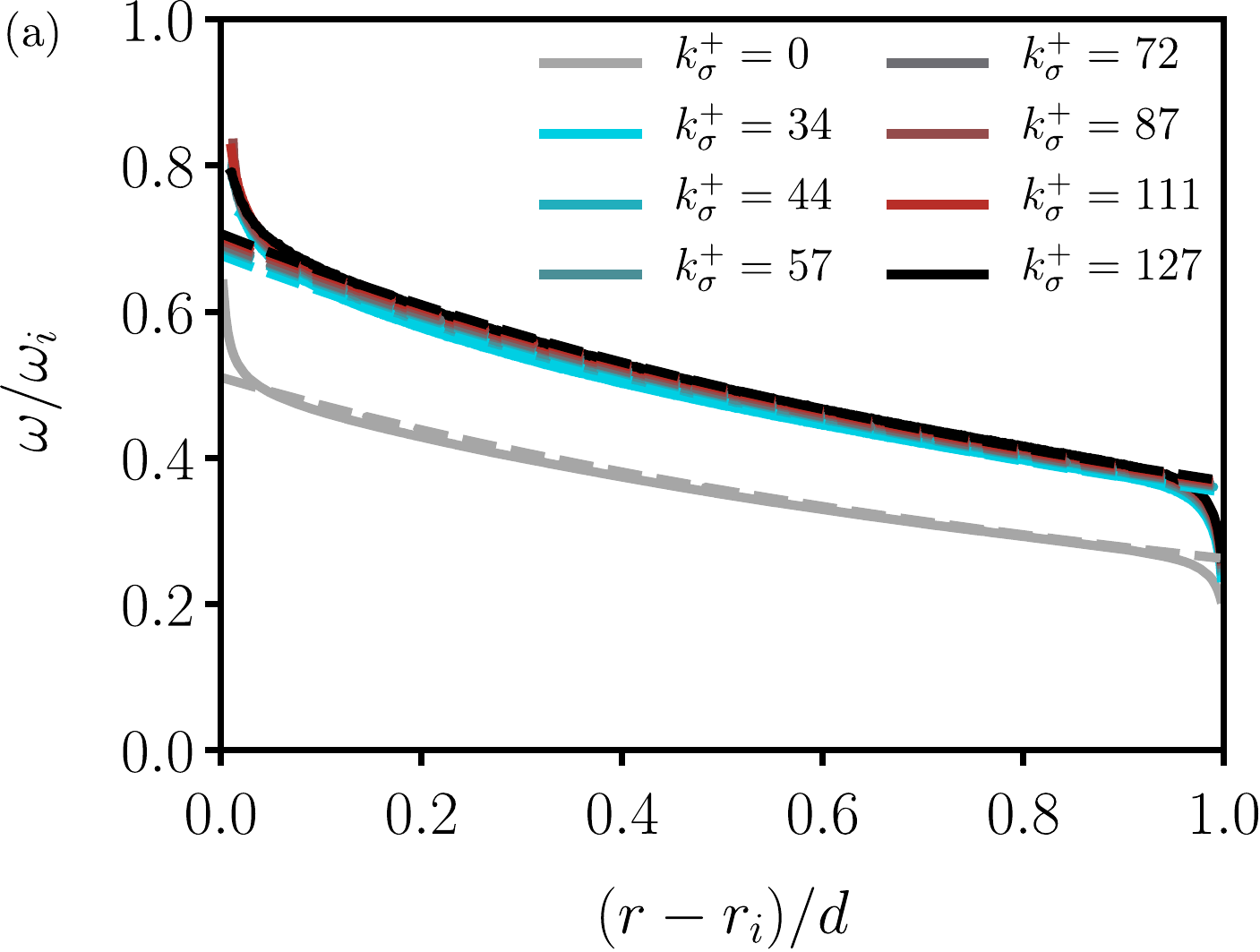}
\end{subfigure}%
\begin{subfigure}{.50\textwidth}
\centering
\includegraphics[width=0.97\linewidth]{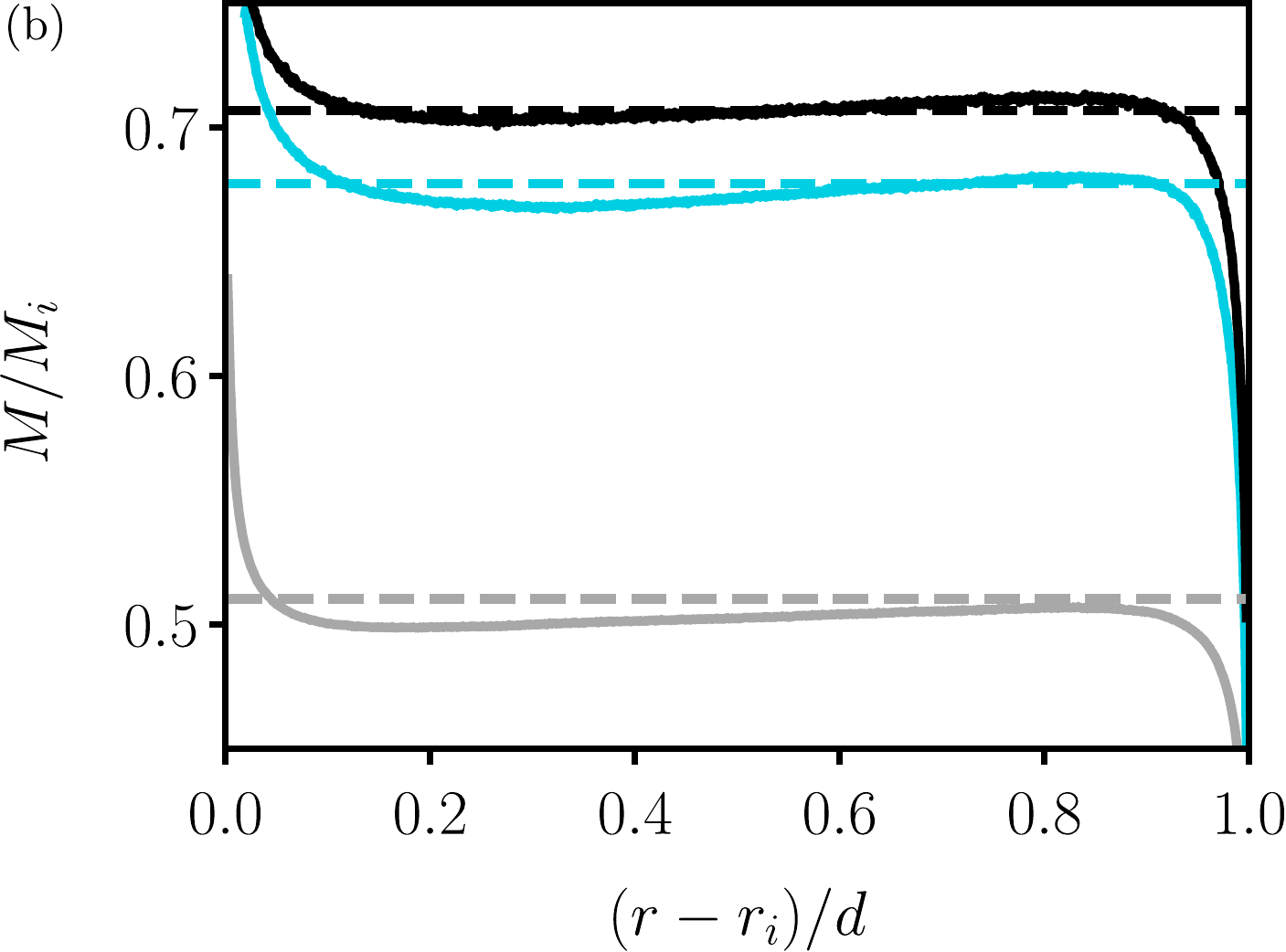}
\end{subfigure}%
\caption{Bulk velocity profiles. (a) The mean angular velocity normalized with the inner velocity $\omega/\omega_i$, versus the radius $(r-r_i)/d$ normalized with the gap width $d$. The profiles for different roughness heights $k_\sigma^+$ are compared. The bulk profile is strongly shifted towards the rough inner cylinder, as the roughness there enhances the coupling between the inner BL and the bulk, similarly as the ribs have done in \cite{zhu18}. (b) The angular momentum $M$, normalized with the inner cylinder angular momentum $M_i$. Solid lines are the PIV results and dashed lines ($M_b/M_i$) are calculated from equation~(\ref{eq:ang_mom}). The colors are the same in both figures. The grey line is the smooth wall profile at $Ta=6.2\times 10^{12}$, obtained from \cite{hui13}. }
\label{fig:angmom}
\end{figure}
Thus far, we have discussed the velocity profiles of the inner cylinder boundary layer, i.e. $y^+<0.65L_c^+$. By means of matching this profile to the bulk velocity profile at boundary layer height, one can derive the relationship between the torque $Nu_\omega(Ta)$ and the velocity of the inner cylinder \citep{che19, ber20b}. For smooth wall inner cylinder rotating turbulent TC flow, it is well known that the angular momentum in the bulk ($M_b$) is constant \citep{wen33, tow76}, and, in fact, very close to half the inner cylinder angular momentum ($M_i=\omega_ir_i^2$), $M_b=0.5M_i$, for stationary outer cylinder.
For rough wall TC flow however, and especially for asymmetric roughness when the inner cylinder is of a different roughness height than the outer cylinder, the exact value of $M_b$ is a priori unknown. However, what was shown is that for very rough walls the bulk azimuthal velocity profile is shifted towards the rough cylinder, due to the stronger coupling to that side thanks to the roughness \citep{zhu17,zhu18,ber19}.

If the bulk region velocity conforms to a constant angular momentum, it should match the angular momentum at the edge of the BL $r=r_i+\delta_r$, where $\delta_r=0.65L_c$. The momentum ratio ($M_b/M_i$) is the angular momentum in the bulk over the angular momentum of the inner cylinder
\begin{equation}\label{eq:ang_mom}
\frac{M_b}{M_i} = \frac{\omega|_{y=\delta_r} (r_i + \delta_r)^2}{\omega_i r_i^2},
\end{equation}
where $\omega|_{y=\delta_r} = \omega_{\tau,i}(\omega_i^+ - \omega_r^+(y^+=\delta_r^+))$, and we use the velocity profile of the rough inner cylinder BL, figure \ref{fig:ill_velprof} and (\ref{eq:smooth_rough})-$\Delta \omega^+$, $\omega_r^+(y^+=\delta_r^+) = \frac{1}{\lambda}\log \delta_r^+ + \left( \frac{1}{\kappa}-\frac{1}{\lambda}\right)\log L_{c,r}^+ + 1.0 - \Delta \omega^+$. Figure \ref{fig:angmom} compares the result from equation~{}(\ref{eq:ang_mom}) (dashed line) with the experimentally obtained velocity profiles (solid lines), demonstrating agreement between the calculated and the measured profiles. This supports the assumption that also the rough-wall velocity profiles conform to a constant angular momentum in the bulk. Finally, we point out that the `overshooting' of the profiles in the bulk, i.e. the slight increase in $M$ with increasing $r$, is likely an effect of the turbulent Taylor vortices, and is therefore expected to depend on the height coordinate $z$ \citep{hui14}. It is due to the detaching plumes which are transported to the other side of the gap by the Taylor rolls. Similar overshooting is well known from temperature profiles in turbulent Rayleigh--B\'enard flow \citep{til93,ahl09}.

\section{Calculation of $Nu_\omega(Ta)$ and $C_f(Re)$}\label{sec:nu_ta}
Since the angular momentum in the bulk is to a good approximation constant, we can match the angular momentum of the inner cylinder BL at BL height with the angular momentum of the outer cylinder BL at BL height, i.e. $M(\delta_{i,r}) = M(\delta_{o,s})$. This approach is based on the matching of BL and bulk velocity profiles in the recent CPS model \citep{che19}. Subscripts (\textit{i},\textit{o}) refer to inner cylinder and outer cylinder BL quantities, where subscripts (\textit{s},\textit{r}) refer to smooth and rough wall quantities, and $\delta=0.65L_c$ for inner cylinder and outer cylinder (rough and smooth) so that $\delta_{i,r}^+=\alpha L_{c,i,r}^+$, $\delta_{o,s}^+=\alpha L_{c,o,s}^+$ with $\alpha = 0.65$. The matching argument becomes
    \begin{equation}
    \label{eq:matchM}
        \left( r_i + \delta_{i,r} \right)^2\omega_{\tau,i} \omega_{IC}^+(\delta_{i,r}^+) = \left( r_o-\delta_{o,s}\right)^2 \omega_{\tau,o}\omega_{OC}^+(\delta_{o,s}^+),
    \end{equation}
where we realize that $\omega_{\tau,o} = \eta^2 \omega_{\tau,i}$. We substitute the BL equations for respectively rough and smooth walls into equation~(\ref{eq:matchM}) and obtain
    \begin{equation}
    \label{eq:mathchM_2}
        \begin{gathered}
        \left( r_i+\delta_{i,r}\right)^2 \omega_{\tau,i} \left( \omega_i^+ -\frac{1}{\lambda} \log(\delta_{i,r}^+) - \left(\frac{1}{\kappa} - \frac{1}{\lambda}\right) \log(L_{c,i,r}^+) - C_i + \Delta \omega^+ \right) =\\
        \left( r_o - \delta_{o,s}\right)^2 \omega_{\tau,o} \left(\frac{1}{\lambda}\log(\delta_{o,s}^+) + \left(\frac{1}{\kappa} - \frac{1}{\lambda}\right) \log(L_{c,o,s}^+) + C_o \right).
        \end{gathered}
    \end{equation}
The rough wall, inner cylinder BL height $\delta_{i,r}^+$, and the velocity shift $\Delta \omega^+$ are functions of the sand grain size $k_s^+$. This makes the matching equation more involved, in comparison to the smooth wall case \citep{che19, ber20b}.

Following \cite{che19}, we now rewrite the equation in terms of $Re_{\tau,i}$ and $Re_i$. The inner cylinder angular velocity becomes
    \begin{equation}\label{eq:os_res}
        \omega_i^+=\frac{Re_i}{2Re_{\tau,i}}.
    \end{equation}
The equivalent sand grand size is
    \begin{equation}\label{eq:ks_rw}
        k_s^+ = 2\frac{k_s}{d}Re_{\tau,i} = \epsilon Re_{\tau,i}.
    \end{equation}
The fully rough asymptote from equation~(\ref{eq:fr_asymptote}) can now be rewritten as
    \begin{equation}\label{eq:domp_rw}
        \Delta \omega^+ = \frac{1}{\kappa}\log (\epsilon Re_{\tau,i}) + A - B.
    \end{equation}
The inner cylinder, rough wall, BL height $\delta_{i,r}^+$ is rewritten from $\delta_{i,r}^+=\alpha L_{c,i,r}^+$ as
    \begin{equation}\label{eq:bl_rw}
        \delta_{i,r}^+ = \frac{2\alpha \eta Re_{\tau,i}}{\kappa (1-\eta) \mathcal{Z}};  \quad \quad \text{with} \quad
        \mathcal{Z} =\left( \frac{Re_i}{2Re_{\tau,i}} + \frac{1}{\kappa}\log (\epsilon Re_{\tau,i}) +A -B \right).
    \end{equation}
The outer cylinder, smooth wall, BL height $\delta_{o,s}^+$ is rewritten from $\delta_{o,r}^+=\alpha L_{c,o,r}^+$ as
    \begin{equation}\label{eq:delta_os}
        \delta_{o,s}^+ = \frac{4\alpha \eta^2 Re_{\tau,i}^2}{\kappa (1-\eta) Re_i},
    \end{equation}
We can now substitute equations (\ref{eq:os_res})--(\ref{eq:delta_os}) into (\ref{eq:mathchM_2}), and obtain
    \begin{equation}
    \label{eq:match_re22}
        \begin{gathered}
        \left(1+\frac{\alpha}{\kappa \mathcal{Z}} \right)^2
        \left( \frac{Re_i}{2Re_{\tau,i}}  -  \frac{1}{\lambda}\log \left( \frac{2\alpha \eta Re_{\tau,i}}{\kappa (1-\eta)\mathcal{Z}}\right) \right. \\
        - \left. \left( \frac{1}{\kappa} - \frac{1}{\lambda} \right)\log \left(\frac{2 \eta Re_{\tau,i}}{\kappa (1-\eta)\mathcal{Z}} \right)+ \frac{1}{\kappa}\log (\epsilon Re_{\tau,i}) +A -B  -C_i \right) = \\
        \left(1-\frac{2 \alpha \eta Re_{\tau,i}}{\kappa Re_i}\right)^2 \left( \frac{1}{\lambda} \log \left( \frac{4\alpha \eta^2 Re_{\tau,i}^2}{\kappa (1-\eta) Re_i}\right) + \left( \frac{1}{\kappa} - \frac{1}{\lambda} \right) \log \left( \frac{4\eta^2 Re_{\tau,i}^2}{\kappa (1-\eta) Re_i}\right) +C_o \right).
        \end{gathered}
    \end{equation}
This implicit equation can be solved numerically to obtain $Re_{\tau,i}(Re_i)$ with parameters $C_i = 1.0, C_o=2.5, A=5.0, B=8.5, \kappa=0.39, \lambda=0.64, \alpha=0.65$ for these experiments, $\eta=0.714$ and $\epsilon=0.9694/80$.
Finally, by means of equations (\ref{eq:Ta}) -- (\ref{eq:Rei}), we express the result $Re_{\tau,i}(Re_i)$ into $Nu_\omega(Ta)$ and $C_f(Re_i)$ respectively.

\begin{figure}
\centering
\begin{subfigure}{.50\textwidth}
  \centering
  \includegraphics[width=1.0\linewidth]{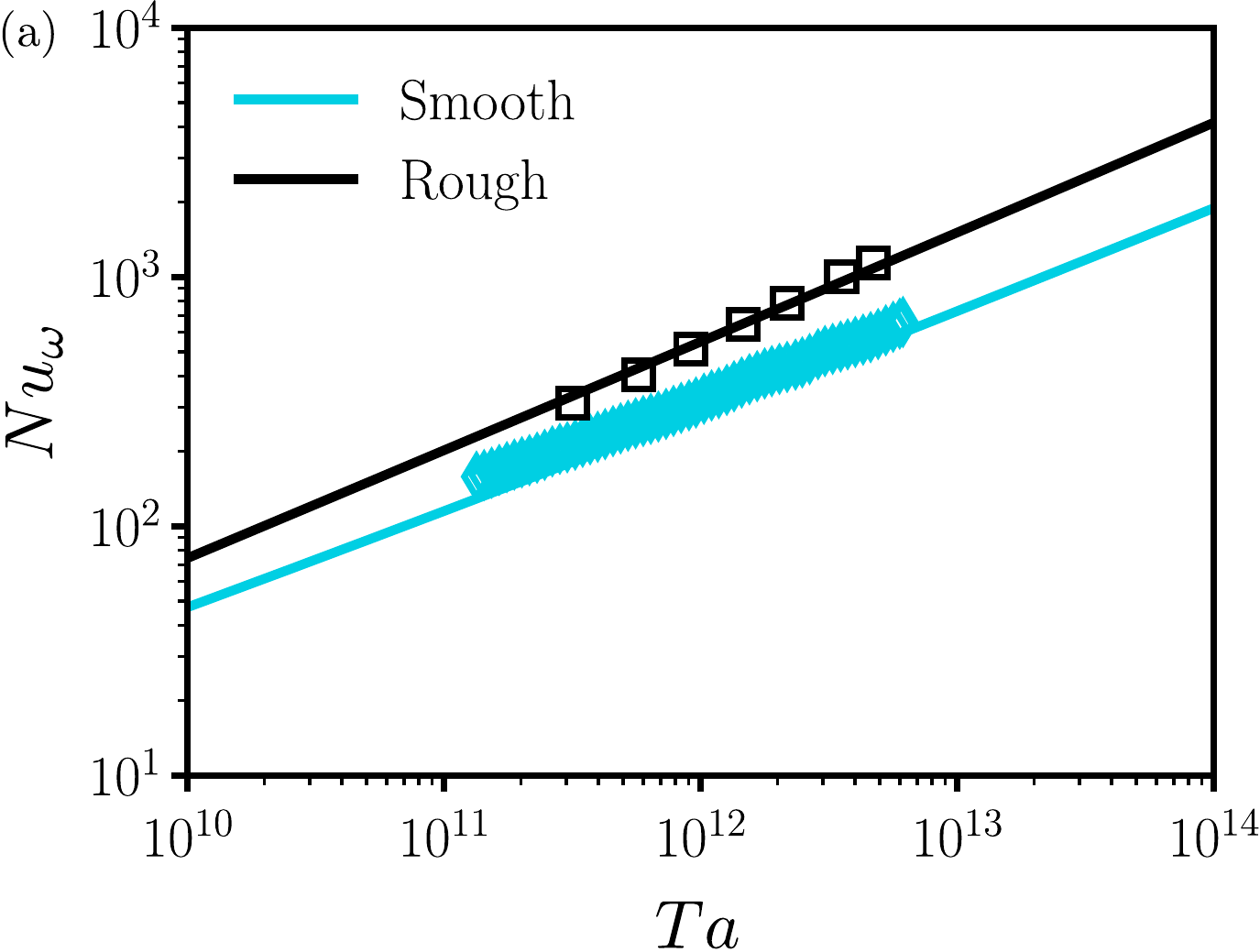}
\end{subfigure}%
\begin{subfigure}{.50\textwidth}
  \centering
  \includegraphics[width=1.0\linewidth]{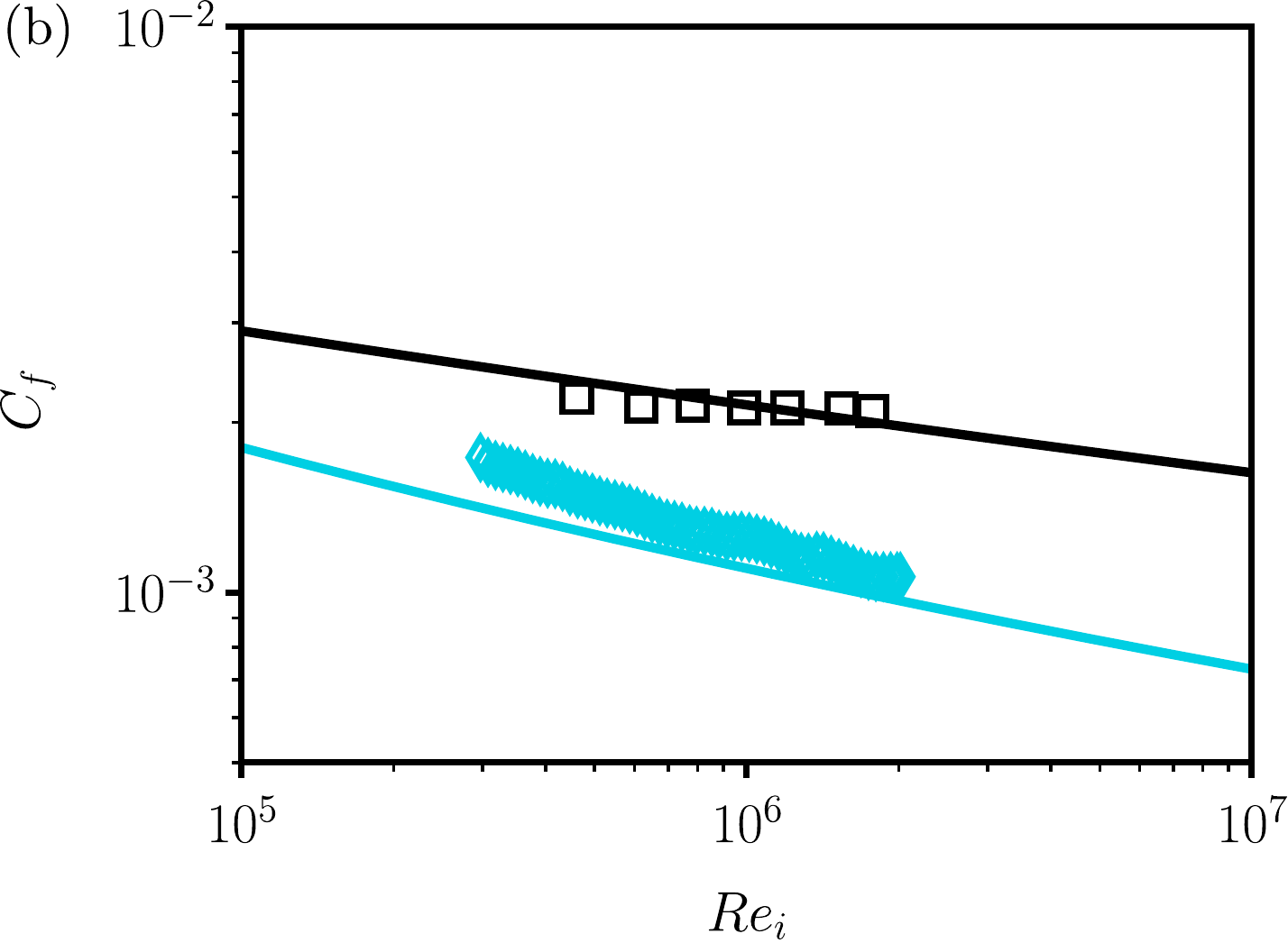}
\end{subfigure}%
    \caption{Global response of the rough $k_s/d=0.012$ and smooth wall TC turbulence. (a) $Nu_\omega = \frac{2Re_\tau^2\eta (1+\eta)}{Re_i}$ versus $Ta$. (b) Friction factor $C_f=\frac{8Re_\tau^2}{Re_i^2}$ versus $Re_i$. Black diamonds are the smooth wall experiments of \cite{gil11}, where the solid black line shows the theory of \cite{ber20b} for smooth wall TC turbulence. Blue squares are the rough inner cylinder measurements from the present work. The solid blue line is equation~(\ref{eq:match_re22}).}
    \label{fig:nuta}
\end{figure}

Figure \ref{fig:nuta} presents the final result, together with the experimental data from smooth walls \citep{gil11} and with the equation for smooth wall TC \citep{ber20b} (grey). The black open squares represent the fully rough inner cylinder rotating TC experiments presently. The black solid line is our calculation from equation~(\ref{eq:match_re22}). We emphasize that no fitting parameters are used. All parameters find their origin in the velocity profiles, and originate from the slopes of the logarithmic velocity profiles ($\kappa^{-1}$, $\lambda^{-1}$), the offset of the smooth velocity profile ($A, C_i, C_o$) or the BL thickness fit for smooth walls $\alpha$. This reflects that all parameters are universal for all radius ratios, and cannot and need not be `tuned'.

The agreement between equation~(\ref{eq:match_re22}) and the experimental data (the maximum error is only $\approx 5\%$) is convincing. It implies that from straightforward measurements of the torque, for given inner cylinder rotation speed, we can calculate the value of $k_s$ with a reasonable accuracy. This means that the TC facility can potentially be used for direct, fast, measurements of surface drag properties, as characterized by $k_s$.

\section{Summary, conclusions, and outlook}\label{sec:sum_con}
We carried out experiments of inner cylinder rotating (and stationary outer cylinder) Taylor--Couette (TC) turbulence with a rough inner cylinder and a smooth outer cylinder. We measured the torque, and, by means of PIV, the mean angular velocity profiles.
The rough surface consisted of P36 industrial grade sandpaper, where the roughness height $k=6k_\sigma$, with $k_\sigma$ the standard deviation of the roughness height, over the gap width $d$ was $k/d=0.014$. The roughness height $k$ was much larger than the viscous length scale $\delta_\nu$, such that $k/\delta_\nu = \numrange{204}{762}$. The velocity shift of the rough wall azimuthal velocity profiles was, compared to the reference smooth wall, in the log-law region $\Delta \omega^+ > 9$ over the whole range of $4.6\times 10^5 < \Rei < 1.77\times 10^6$. Hence the sandpaper was hydrodynamically fully rough. Furthermore, the roughness height $k_\sigma<0.015L_c$, where $0.20L_c$ is the height \citep{ber20b} which separates the region in the BL where production of turbulence is dominated by shear, and the region in the BL where production of turbulence is affected by effects of streamwise curvature.

Using the mean azimuthal velocity profiles, we found that the slope of the fully rough asymptote, characterized by $\kappa= 0.34\pm 0.02$, was similar to previous findings in flat plate BLs $\kappa \approx 0.38$. Also, the value of the equivalent sand grain roughness height $k_s$ compared reasonably well with those found for sandpaper in flat plate BLs \citep{fla07, squ16}.

Finally, to obtain the relationship between the dimensionless torque and dimensionless driving of the system $Nu_\omega(Ta)$, we employed a matching argument between the inner cylinder BL rough mean angular momentum profile at the inner cylinder BL height, and the smooth outer cylinder BL mean angular momentum profile, at the outer cylinder BL height, based on the CPS model of \cite{che19}, see also \cite{ber20b}. To justify this, we first showed that for a rough wall inner cylinder, a region of constant angular momentum exists in the bulk. We find a convincing overlap between the calculated value of the torque (or wall shear stress), and the experimentally measured values of the torque, with a maximum error of $\approx 5\%$.

These findings indicate that the turbulent TC facility can be a valuable setup for characterizing the turbulent drag properties of any rough surface. Direct and straightforward measurements of the torque can now be translated to a value of the equivalent sand grain roughness height $k_s$. It seems that the value of $k_s$ found in TC is similar to the value of $k_s$ found in flat plate BLs.

As an outlook to future work, we propose that more studies in both turbulent flat plate BLs
and turbulent TC flow, with identical rough surfaces, are carried out to further compare
the drag properties of these surfaces. Further unanswered questions include the effects of even more considerable roughness penetrating the curvature affected logarithmic regime of the BL, which is related to finding the slope of the fully rough asymptote in that region. This could also be achieved by employing a TC setup with a lower radius ratio $\eta$, thus increasing curvature effects.

\section*{Acknowledgements}
We would like to thank B. Benschop, M. Bos, and G.W. Bruggert for their technical support, Y.A. Lee for his support in the lab and M.A. Bruning for discussions.

This study was funded by the Netherlands Organisation for Scientific Research (NWO) through the Multiscale Catalytic Energy Conversion (MCEC) research center and the GasDrive project 14504, by the European Research Council (ERC) Advanced Grant ``Droplet Diffusive Dynamics'', and by the Priority Programme SPP 1881 Turbulent Superstructures of the Deutsche Forschungs-gemeinschaft.\\

P.B. and P.A.B. are shared first author.



\bibliographystyle{jfm}
\bibliography{literature_turbulence.bib}

\begin{thebibliography}{45}
\expandafter\ifx\csname natexlab\endcsname\relax\def\natexlab#1{#1}\fi
\def\au#1{#1} \def\ed#1{#1} \def\yr#1{#1}\def\at#1{#1}\def\jt#1{\textit{#1}}
  \def\bt#1{#1}\def\bvol#1{\textbf{#1}} \def\vol#1{#1} \def\pg#1{#1}
  \def\publ#1{#1}\def\arxiv#1{#1}\def\org#1{#1}\def\st#1{\textit{#1}}

\bibitem[Ahlers {\em et~al.\/}(2009)Ahlers, Grossmann \& Lohse]{ahl09}
{\sc \au{Ahlers, G.}, \au{Grossmann, S.} \& \au{Lohse, D.}} \yr{2009}  \at{Heat
  transfer and large scale dynamics in turbulent {{Rayleigh-B\'enard}}
  convection}.  \jt{Rev. Mod. Phys.}  \bvol{81},  \pg{503}.

\bibitem[Baars {\em et~al.\/}(2016)Baars, Squire, Talluru, Abbassi, Hutchins \&
  Marusic]{baa16}
{\sc \au{Baars, W.~J.}, \au{Squire, D.~T.}, \au{Talluru, K.~M.}, \au{Abbassi,
  M.~R.}, \au{Hutchins, N.} \& \au{Marusic, I.}} \yr{2016}  \at{Wall-drag
  measurements of smooth- and rough-wall turbulent boundary layers using a
  floating element}.  \jt{Exp. Fluids}  \bvol{57}~(90).

\bibitem[Bakhuis {\em et~al.\/}(2020)Bakhuis, Ezeta, Berghout, Bullee, Tai,
  Chung, Verzicco, Lohse, Huisman \& Sun]{bak20}
{\sc \au{Bakhuis, D.}, \au{Ezeta, R.}, \au{Berghout, P.}, \au{Bullee, P.A.},
  \au{Tai, N.C.}, \au{Chung, D.}, \au{Verzicco, R.}, \au{Lohse, D.},
  \au{Huisman, S.G.} \& \au{Sun, C.}} \yr{2020}  \at{Controlling secondary flow
  in {Taylor--Couette} turbulence through spanwise-varying roughness}.  \jt{J.
  Fluid Mech.}  \bvol{883},  \pg{A15}.

\bibitem[van~den Berg {\em et~al.\/}(2003)van~den Berg, Doering, Lohse \&
  Lathrop]{ber03}
{\sc \au{van~den Berg, T.~H.}, \au{Doering, C.~R.}, \au{Lohse, D.} \&
  \au{Lathrop, D.}} \yr{2003}  \at{Smooth and rough boundaries in turbulent
  {{Taylor-Couette}} flow}.  \jt{Phys. Rev. E}  \bvol{68},  \pg{036307}.

\bibitem[Berghout {\em et~al.\/}(2020)Berghout, Verzicco, Stevens, Lohse \&
  Chung]{ber20b}
{\sc \au{Berghout, P.}, \au{Verzicco, R.}, \au{Stevens, R. J. A.~M.},
  \au{Lohse, D.} \& \au{Chung, D.}} \yr{2020}  \at{Calculation of the mean
  velocity profile for strongly turbulent {Taylor--Couette} flow and arbitrary
  radius ratios}.  \jt{J. Fluid Mech.}  \bvol{In press}.

\bibitem[Berghout {\em et~al.\/}(2019)Berghout, Zhu, Chung, Verzicco, Stevens
  \& Lohse]{ber19}
{\sc \au{Berghout, P.}, \au{Zhu, X.}, \au{Chung, D.}, \au{Verzicco, R.},
  \au{Stevens, R.J.A.M.} \& \au{Lohse, D.}} \yr{2019}  \at{Direct numerical
  simulations of {Taylor--Couette} turbulence: the effects of sand grain
  roughness}.  \jt{J. Fluid Mech.}  \bvol{873},  \pg{260--286}.

\bibitem[Bradshaw(1969)]{bra69}
{\sc \au{Bradshaw, P.}} \yr{1969}  \at{The analogy between streamline curvature
  and buoyancy in turbulent shear flow}.  \jt{J. Fluid Mech.}  \bvol{36},
  \pg{177--191}.

\bibitem[Cadot {\em et~al.\/}(1997)Cadot, Couder, Daerr, Douady \&
  Tsinober]{cad97}
{\sc \au{Cadot, O.}, \au{Couder, Y.}, \au{Daerr, A.}, \au{Douady, S.} \&
  \au{Tsinober, A.}} \yr{1997}  \at{Energy injection in closed turbulent flows:
  Stirring through boundary layers versus inertial stirring}.  \jt{Phys. Rev.
  E}  \bvol{56},  \pg{427--433}.

\bibitem[Cheng {\em et~al.\/}(2020)Cheng, Pullin \& Samtaney]{che19}
{\sc \au{Cheng, W.}, \au{Pullin, D.~I.} \& \au{Samtaney, R.}} \yr{2020}
  \at{{Large--eddy simulation and modeling of Taylor--Couette flow with an
  outer stationary cylinder}}.  \jt{J. Fluid Mech.}  \bvol{890},  \pg{A17}.

\bibitem[Chung {\em et~al.\/}(2021)Chung, Hutchins, Schultz \& Flack]{chu21}
{\sc \au{Chung, D.}, \au{Hutchins, N.}, \au{Schultz, M.~P.} \& \au{Flack,
  K.~A.}} \yr{2021}  \at{Predicting the drag of roughness}.  \jt{Annu. Rev.
  Fluid Mech.}  \bvol{In press}.

\bibitem[Clauser(1954)]{cla54}
{\sc \au{Clauser, F.~H.}} \yr{1954}  \at{Turbulent boundary layers in adverse
  pressure gradients.}  \jt{J. Aeronaut. Sci.}  \bvol{21},  \pg{91--108}.

\bibitem[Eckhardt {\em et~al.\/}(2007)Eckhardt, Grossmann \& Lohse]{eck07b}
{\sc \au{Eckhardt, B.}, \au{Grossmann, S.} \& \au{Lohse, D.}} \yr{2007}
  \at{Torque scaling in turbulent {{Taylor-Couette}} flow between independently
  rotating cylinders}.  \jt{J. Fluid Mech.}  \bvol{581},  \pg{221--250}.

\bibitem[Flack {\em et~al.\/}(2020)Flack, Schultz \& J.M.]{fla20}
{\sc \au{Flack, K.A.}, \au{Schultz, M.P.} \& \au{J.M., Barros}} \yr{2020}
  \at{Skin friction measurements of systematically-varied roughness: Probing
  the role of roughness amplitude and skewness}.  \jt{Flow Turb. Combust.}
  \bvol{104},  \pg{317–329}.

\bibitem[Flack \& Schultz(2010)]{fla10}
{\sc \au{Flack, K.~A.} \& \au{Schultz, M.~P.}} \yr{2010}  \at{Review of
  hydraulic roughness scales in the fully rough regime}.  \jt{Trans. ASME: J.
  Fluids Eng.}  \bvol{132},  \pg{041203}.

\bibitem[Flack {\em et~al.\/}(2007)Flack, Schultz \& Connelly]{fla07}
{\sc \au{Flack, K.~A.}, \au{Schultz, M.~P.} \& \au{Connelly, J.~S.}} \yr{2007}
  \at{Examination of a critical roughness height for outer layer similarity}.
  \jt{Physics of Fluids}  \bvol{19}~(9),  \pg{095104}.

\bibitem[Forooghi {\em et~al.\/}(2017)Forooghi, Stroh, Magagnato, Jakirli\'c \&
  Frohnapfel]{for17}
{\sc \au{Forooghi, P.}, \au{Stroh, A.}, \au{Magagnato, F.}, \au{Jakirli\'c, S.}
  \& \au{Frohnapfel, B.}} \yr{2017}  \at{Toward a universal roughness
  correlation}.  \jt{J. Fluids Eng.}  \bvol{139},  \pg{121201}.

\bibitem[van Gils {\em et~al.\/}(2011{\natexlab{{\em a\/}}})van Gils, Bruggert,
  Lathrop, Sun \& Lohse]{gil11a}
{\sc \au{van Gils, D. P.~M.}, \au{Bruggert, G.~W.}, \au{Lathrop, D.~P.},
  \au{Sun, C.} \& \au{Lohse, D.}} \yr{2011{\natexlab{{\em a\/}}}}  \at{The
  {{Twente}} turbulent {{Taylor-Couette}} {{($T^3C$)}} facility: strongly
  turbulent (multi-phase) flow between independently rotating cylinders}.
  \jt{Rev. Sci. Instrum.}  \bvol{82},  \pg{025105}.

\bibitem[van Gils {\em et~al.\/}(2011{\natexlab{{\em b\/}}})van Gils, Huisman,
  Bruggert, Sun \& Lohse]{gil11}
{\sc \au{van Gils, D. P.~M.}, \au{Huisman, S.~G.}, \au{Bruggert, G.~W.},
  \au{Sun, C.} \& \au{Lohse, D.}} \yr{2011{\natexlab{{\em b\/}}}}  \at{Torque
  scaling in turbulent {{Taylor-Couette}} flow with co- and counter-rotating
  cylinders}.  \jt{Phys. Rev. Lett.}  \bvol{106},  \pg{024502}.

\bibitem[van Gils {\em et~al.\/}(2012)van Gils, Huisman, Grossmann, Sun \&
  Lohse]{gil12}
{\sc \au{van Gils, D. P.~M.}, \au{Huisman, S.~G.}, \au{Grossmann, S.}, \au{Sun,
  C.} \& \au{Lohse, D.}} \yr{2012}  \at{Optimal {{Taylor-Couette}} turbulence}.
   \jt{J. Fluid Mech.}  \bvol{706},  \pg{118--149}.

\bibitem[Grossmann {\em et~al.\/}(2014)Grossmann, Lohse \& Sun]{gro14}
{\sc \au{Grossmann, S.}, \au{Lohse, D.} \& \au{Sun, C.}} \yr{2014}
  \at{Velocity profiles in strongly turbulent {{Taylor-Couette}} flow}.
  \jt{Phys. Fluids}  \bvol{26},  \pg{025114}.

\bibitem[Grossmann {\em et~al.\/}(2016)Grossmann, Lohse \& Sun]{gro16}
{\sc \au{Grossmann, S.}, \au{Lohse, D.} \& \au{Sun, C.}} \yr{2016}  \at{High
  {{Reynolds}} number {{Taylor-Couette}} turbulence}.  \jt{Annu. Rev. Fluid
  Mech.}  \bvol{48},  \pg{53--80}.

\bibitem[Hama(1954)]{ham54}
{\sc \au{Hama, F.R.}} \yr{1954}  \at{Boundary-layer characteristics for smooth
  and rough surfaces.}  \jt{Trans. Soc. Nav. Archit. Mar. Engrs}  \bvol{62},
  \pg{333--358}.

\bibitem[Huisman {\em et~al.\/}(2013)Huisman, Scharnowski, Cierpka, K{\"a}hler,
  Lohse \& Sun]{hui13}
{\sc \au{Huisman, S.~G.}, \au{Scharnowski, S.}, \au{Cierpka, C.},
  \au{K{\"a}hler, C.~J.}, \au{Lohse, D.} \& \au{Sun, C.}} \yr{2013}
  \at{Logarithmic boundary layers in strong {{Taylor-Couette}} turbulence}.
  \jt{Phys. Rev. Lett.}  \bvol{110},  \pg{264501}.

\bibitem[Huisman {\em et~al.\/}(2014)Huisman, van~der Veen, Sun \&
  Lohse]{hui14}
{\sc \au{Huisman, S.~G.}, \au{van~der Veen, R. C.~A.}, \au{Sun, C.} \&
  \au{Lohse, D.}} \yr{2014}  \at{Multiple states in highly turbulent
  {{Taylor-Couette}} flow}.  \jt{Nature Commun.}  \bvol{5},  \pg{3820}.

\bibitem[Jim{\'e}nez(2004)]{jim04}
{\sc \au{Jim{\'e}nez, J.}} \yr{2004}  \at{Turbulent flows over rough walls}.
  \jt{Annu. Rev. Fluid Mech.}  \bvol{36},  \pg{173--196}.

\bibitem[K\"ahler {\em et~al.\/}(2012)K\"ahler, Scharnowski \& Cierpka]{kah12a}
{\sc \au{K\"ahler, C.~J.}, \au{Scharnowski, S.} \& \au{Cierpka, C.}} \yr{2012}
  \at{On the resolution limit of digital particle image velocimetry}.  \jt{Exp.
  Fluids}  \bvol{52},  \pg{1629--1639}.

\bibitem[K\"ahler {\em et~al.\/}(2006)K\"ahler, Scholz \& Ortmanns]{kahler2006}
{\sc \au{K\"ahler, C.~J.}, \au{Scholz, U.} \& \au{Ortmanns, J.}} \yr{2006}
  \at{Wall-shear-stress and near-wall turbulence measurements up to single
  pixel resolution by means of long-distance micro-piv}.  \jt{Experiments in
  Fluids}  \bvol{41},  \pg{327--341}.

\bibitem[Lathrop {\em et~al.\/}(1992)Lathrop, Fineberg \& Swinney]{lat92a}
{\sc \au{Lathrop, D.~P.}, \au{Fineberg, J.} \& \au{Swinney, H.~S.}} \yr{1992}
  \at{Transition to shear-driven turbulence in {{Couette-Taylor}} flow}.
  \jt{Phys. Rev. A}  \bvol{46},  \pg{6390--6405}.

\bibitem[Monin \& Yaglom(1975)]{mon75}
{\sc \au{Monin, A.~S.} \& \au{Yaglom, A.~M.}} \yr{1975} {\em Statistical Fluid
  Mechanics\/}.  \publ{Cambridge, Mass.: MIT Press}.

\bibitem[Napoli {\em et~al.\/}(2008)Napoli, Armenio \& Marchis]{nap08}
{\sc \au{Napoli, E.}, \au{Armenio, V.} \& \au{Marchis, M.~De}} \yr{2008}
  \at{The effect of the slope of irregularly distributed roughness elements on
  turbulent wall-bounded flows}.  \jt{J. Fluid Mech.}  \bvol{613},
  \pg{385--394}.

\bibitem[Nikuradse(1933)]{nik33}
{\sc \au{Nikuradse, J.}} \yr{1933}  \at{{Str\"{o}mungsgesetze} in rauhen
  {Rohren} (trans. {\it flow laws in rough pipes})}.  \jt{Forschungsheft Arb.
  Ing.-Wes.}  \bvol{361}.

\bibitem[Obukhov(1971)]{obu71}
{\sc \au{Obukhov, A.~M.}} \yr{1971}  \at{Turbulence in an atmosphere with a
  non-uniform temperature}.  \jt{Bound.-Layer Meteorol.}  \bvol{2},
  \pg{7--29}.

\bibitem[Ostilla-M\'onico {\em et~al.\/}(2015)Ostilla-M\'onico, Verzicco,
  Grossmann \& Lohse]{ost15b}
{\sc \au{Ostilla-M\'onico, R.}, \au{Verzicco, R.}, \au{Grossmann, S.} \&
  \au{Lohse, D.}} \yr{2015}  \at{The near-wall region of highly turbulent
  {{Taylor-Couette}} flow}.  \jt{J. Fluid Mech.}  \bvol{788},  \pg{95--117}.

\bibitem[Pope(2000)]{pop00}
{\sc \au{Pope, S.~B.}} \yr{2000} {\em Turbulent Flow\/}.  \publ{Cambridge:
  Cambridge University Press}.

\bibitem[Raupach {\em et~al.\/}(1991)Raupach, Antonia \& Rajagopalan]{rau91}
{\sc \au{Raupach, M.R.}, \au{Antonia, R.A.} \& \au{Rajagopalan, S.S.}}
  \yr{1991}  \at{Rough-wall turbulent boundary layers.}  \jt{J. Fluids Eng.}
  \bvol{44}~(1),  \pg{1--25}.

\bibitem[Scharnowski {\em et~al.\/}(2012)Scharnowski, Hain \& K\"ahler]{sch12b}
{\sc \au{Scharnowski, S.}, \au{Hain, R.} \& \au{K\"ahler, C.~J.}} \yr{2012}
  \at{Reynolds stress estimation up to single-pixel resolution using
  piv-measurements}.  \jt{Exp. Fluids}  \bvol{52},  \pg{985--1002}.

\bibitem[Squire {\em et~al.\/}(2016)Squire, Morrill-Winter, Hutchins, Schultz,
  Klewicki \& Marusic]{squ16}
{\sc \au{Squire, D.~T.}, \au{Morrill-Winter, C.}, \au{Hutchins, N.},
  \au{Schultz, M.~P.}, \au{Klewicki, J.~C.} \& \au{Marusic, I.}} \yr{2016}
  \at{Comparison of turbulent boundary layers over smooth and rough surfaces up
  to high {{Reynolds}} numbers}.  \jt{J. Fluid Mech.}  \bvol{795},
  \pg{210--240}.

\bibitem[Taylor(1923)]{tay23}
{\sc \au{Taylor, G.~I.}} \yr{1923}  \at{Experiments on the motion of solid
  bodies in rotating fluids}.  \jt{Proc. R. Soc. London A}  \bvol{104},
  \pg{213--218}.

\bibitem[Tilgner {\em et~al.\/}(1993)Tilgner, Belmonte \& Libchaber]{til93}
{\sc \au{Tilgner, A.}, \au{Belmonte, A.} \& \au{Libchaber, A.}} \yr{1993}
  \at{Temperature and velocity profiles of turbulence convection in water}.
  \jt{Phys. Rev. E}  \bvol{47},  \pg{R2253--R2256}.

\bibitem[Townsend(1976)]{tow76}
{\sc \au{Townsend, A.~A.}} \yr{1976} {\em The Structure of Turbulent Shear
  Flow, 2st edn.\/}.  \publ{Cambridge, UK: Cambridge University Press}.

\bibitem[van~der Veen {\em et~al.\/}(2016)van~der Veen, Huisman, Merbold,
  Harlander, Egbers, Lohse \& Sun]{vee16}
{\sc \au{van~der Veen, R. C.~A.}, \au{Huisman, S.~G.}, \au{Merbold, S.},
  \au{Harlander, W.}, \au{Egbers, C.}, \au{Lohse, D.} \& \au{Sun, C.}}
  \yr{2016}  \at{{{Taylor-Couette}} turbulence at radius ratio $\eta=0.5$:
  scaling, flow structures and plumes}.  \jt{J. Fluid Mech.}  \bvol{799},
  \pg{334--351}.

\bibitem[Verschoof {\em et~al.\/}(2018)Verschoof, Zhu, Bakhuis, Huisman,
  Verzicco, Sun \& Lohse]{ver18}
{\sc \au{Verschoof, R.~A.}, \au{Zhu, X.}, \au{Bakhuis, D.}, \au{Huisman,
  S.~G.}, \au{Verzicco, R.}, \au{Sun, C.} \& \au{Lohse, D.}} \yr{2018}
  \at{Rough-wall turbulent taylor-couette flow: The effect of the rib height}.
  \jt{Eur. Phys. J. E.}  \bvol{41}~(10),  \pg{125}.

\bibitem[Wendt(1933)]{wen33}
{\sc \au{Wendt, F.}} \yr{1933}  \at{Turbulente {{Str{\"o}mungen}} zwischen zwei
  rotierenden {{Zylindern}}}.  \jt{Ingenieurs-Archiv}  \bvol{4},
  \pg{577--595}.

\bibitem[Zhu {\em et~al.\/}(2018)Zhu, Verschoof, Bakhuis, Huisman, Verzicco,
  Sun \& Lohse]{zhu18}
{\sc \au{Zhu, X.}, \au{Verschoof, R.~A.}, \au{Bakhuis, D.}, \au{Huisman,
  S.~G.}, \au{Verzicco, R.}, \au{Sun, C.} \& \au{Lohse, D.}} \yr{2018}
  \at{Wall roughness induces asymptotic ultimate turbulence}.  \jt{Nature
  Physics}  \bvol{14},  \pg{417--423}.

\bibitem[Zhu {\em et~al.\/}(2017)Zhu, Verzicco \& Lohse]{zhu17}
{\sc \au{Zhu, X.}, \au{Verzicco, R.} \& \au{Lohse, D.}} \yr{2017}
  \at{Disentangling the origins of torque enhancement through wall roughness in
  {{Taylor-Couette}} turbulence}.  \jt{J. Fluid Mech.}  \bvol{812},
  \pg{279--293}.

\end{thebibliography}

\end{document}